\documentclass[runningheads]{llncs}
\usepackage{graphicx}

\usepackage{tikz}
\usepackage{comment}
\usepackage{amsmath,amssymb} 
\usepackage{color}

\usepackage[accsupp]{axessibility}  

\usepackage{amsmath}
\usepackage{amsfonts}
\usepackage{booktabs}
\usepackage{graphicx}
\usepackage{cite}
\usepackage{caption}
\usepackage{subcaption}
\usepackage[ruled,linesnumbered]{algorithm2e}
\usepackage{adjustbox}
\usepackage{hyperref}

\newcommand{\etal}{{\em et al.}~}
\newcommand{\ie}{{\em i.e.,}~}
\newcommand{\eg}{{\em e.g.,}~}

\newcommand{\gf}[2]{{\color{gray}#1}{\color{purple}#2}}

\newcommand{\ninf}[2]{\|\nabla_{#1} #2\|_\infty}

\DeclareMathOperator*{\argmin}{argmin}

\begin{document}

\pagestyle{headings}
\mainmatter
\def\ECCVSubNumber{2877}  

\title{Fast Two-step Blind Optical\\ Aberration Correction}

\titlerunning{Fast Two-step Blind Optical Aberration Correction}
\author{Thomas Eboli \and
Jean-Michel Morel \and
Gabriele Facciolo}
\authorrunning{T. Eboli et al.}
\institute{Universit\'e Paris-Saclay, ENS Paris-Saclay,  CNRS, Centre Borelli, France\\
\url{https://github.com/teboli/fast_two_stage_psf_correction}}
\maketitle

\begin{abstract}
The optics of any camera degrades the sharpness of photographs, which is a key visual quality criterion. This degradation is characterized by the point-spread function (PSF), which depends on the wavelengths of light and is variable across the imaging field.
In this paper, we propose a two-step scheme to correct optical aberrations 
in a single raw or JPEG image, 
\ie without any prior information on the camera or lens.
First, we estimate local Gaussian blur kernels for overlapping patches and sharpen them with a non-blind deblurring technique.
Based on the measurements of the PSFs of dozens of lenses, these blur kernels are modeled as RGB Gaussians defined by seven parameters.
Second, we remove the remaining lateral chromatic aberrations (not contemplated in the first step) 
with a convolutional neural network, trained to
minimize the red/green and blue/green residual images.
Experiments on both synthetic and real images show that
the combination of these two stages yields
a fast state-of-the-art blind optical aberration compensation technique that competes with commercial non-blind algorithms.

\end{abstract}

\section{Introduction}

\begin{figure}[t]
    \centering
    \includegraphics[width=\textwidth]{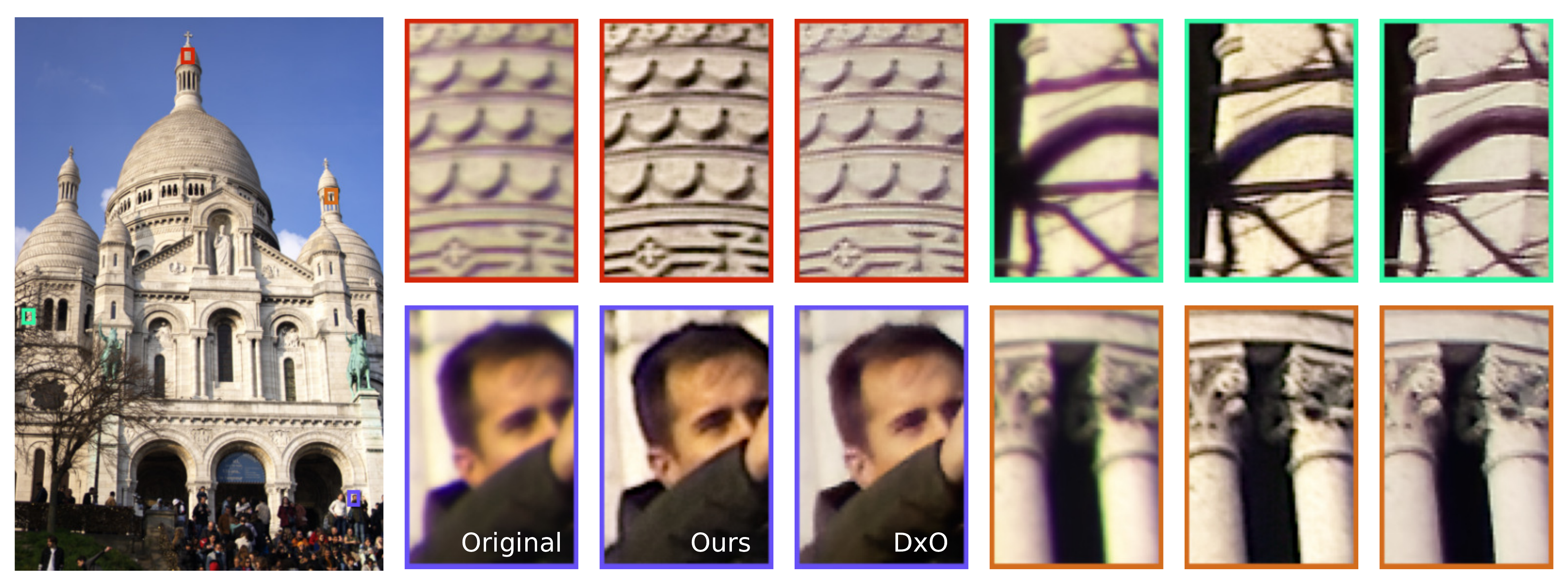}
    
    \caption{We propose a blind method to correct the optical
    aberrations caused by the point-spread function of the lens, without 
    any prior on the lens or the camera to restore the image.
    We sharpen and compensate the visible colored fringes in 
    a 24 megapixels ($4000\times6000$) photograph 
    taken with a Sony $\alpha6000$ camera
    and a Sony FE 35mm $f/1.8$ lens at maximal aperture in 2 seconds
    on a NVIDIA 3090 GPU, 
    achieving a visual result comparable to that of the
    non-blind algorithm of DxO PhotoLab (best seen on a computer screen).}
    \label{fig:teaser}
\end{figure}

Sharpness is a critical criterion for both  photographers
and scientific applications. 
In the absence of motion and with perfect focus, there will
always be blur in the raw photographs, caused by the optics.
The choice of the objective is thus important to take the best possible images and its quality is often characterized by its point spread
function (or {\em PSF}), which is the combination of the optical aberrations 
transforming a white point in the ideal focal image into a colored spot.
In real images, the PSF introduces optical aberrations degrading the global sharpness and introducing colored fringes next to the contrasted edges, \gf{}{see} for instance in Fig.~\ref{fig:teaser} for a mid-entry camera/lens pair.

\begin{figure}[t]
    \centering
    \includegraphics[width=0.90\textwidth]{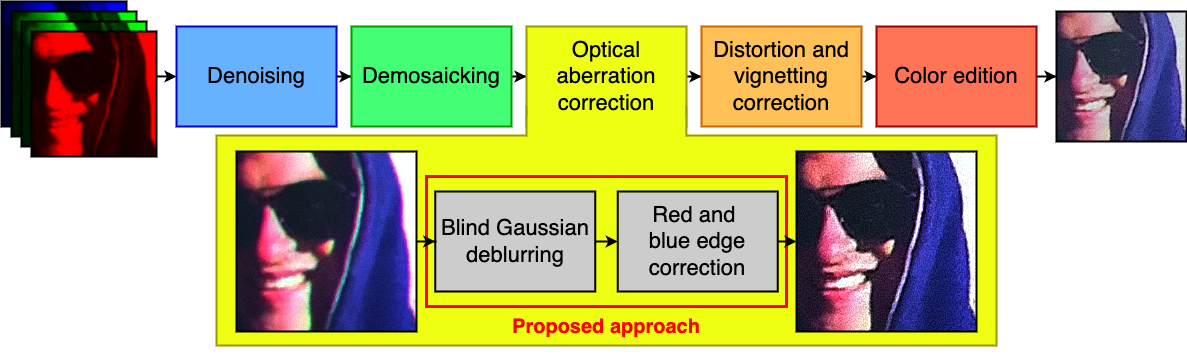}
  
    \caption{Main stages of an editing software,
    processing a raw photograph into a JPEG image. We focus on the
    optical aberration correction module, usually just after denoising
    and demosaicking and before further 
    color and geometry corrections. We decompose this block into two stages: (i) we improve sharpness with a blind deblurring algorithm, and (ii) 
    we align the contrasted red and blue edges to remove the 
    colored fringes at the vicinity of contrasted edges.}
    \label{fig:editingstages}
\end{figure}

Since most cameras use glass or plastic lenses, the effects of the PSF 
cannot be avoided but only compensated 
by either switching to a better objective with a smaller colored spot,
or post-processing the aberrated photographs.
The first solution seems to be the most appealing 
since it solves the problem at its root
but the top-of-the-line objectives are too expensive for most consumers.
Furthermore, most pictures are taken nowadays with smartphone cameras
that have low-quality and non-interchangeable lenses, hence the relevance of efficient algorithmic solutions.
Optical aberration correction, along with denoising, demosaicking and 
distortion and vignetting correction, is among the earliest processing steps of any commercial editing software, 
\eg Adobe Lightroom or DxO PhotoLab.
Figure~\ref{fig:editingstages} shows an example of such
an image processing pipeline.
However, these software rely on accurate calibration of camera/objective pairs,
which are based on exhaustive measurements of all 
the possible camera settings.

In this paper, we propose a blind optical aberration
compensation technique that can be applied to any raw or JPEG image {\em without}
any prior knowledge of the camera or lens. 
Unlike the current state of the art that casts this problem correction as an instance of blind 
deblurring with RGB kernels~\cite{schuler11nonstationnary,yue15geometric,li21universal}, we follow~\cite{kang07automatic}
and decompose optical aberration compensation 
into a two-stage scheme that first removes lens blur and second
compensates the remaining color fringes.
We show a visual comparison with the non-blind commercial solution of DxO
in Figure~\ref{fig:teaser}.
Our deblurring stage relies on the observation that the real RGB PSF measurements of \cite{bauer18mtf} and the parametric kernels of \cite{kee11modeling} (which model local RGB kernels of real data),  fit 2D Gaussian filters defined by just seven parameters.
We confirm that these Gaussian filters verify the ``mild blur'' condition needed to apply the fast blind deblurring algorithm proposed in~\cite{delbracio21polyblur}.
We thus adapt this approach to our problem to increase the sharpness
of overlapping patches, assuming the blur is uniform on their supports. 
We correct the remaining effects due to the color-dependent warp by independently processing 
the red and blue channels using a small convolutional neural network (CNN) trained to minimize the red/green and blue/green image residuals.
This is motivated by the analysis of color fringes in~\cite{chang13correction} showing that
the profile of this image transformation is directly related to the intensity of the colored fringes.
Thanks to the above decomposition, a shallow 160K-parameter CNN is enough
to achieve state-of-the-art results.
We finally gather the patches processed by the CNN.

Our approach presents several advantages over concurrent academic 
works and commercial solutions.
First, the blind deblurring stage is very fast and memory-efficient 
since it leverages the Gaussian model of \cite{kee11modeling}
and the approximated deconvolution scheme from \cite{delbracio21polyblur}. 
Moreover, since our 2D Gaussian lens blur approximation only has a seven  parameters, it is easy to
compute. Yet, the method yields satisfactory visual results.
Second, our approach does not suppose any parametric warp model to represent
the displacements of the edges in the red and blue channels,
which results in a more accurate prediction and in a method that may run
either on crops or the full image. Furthermore, since the colored fringes
are relatively thin, a small, fast and memory-efficient CNN architecture
yields satisfactory results.
Third, since the method is blind to the camera and lens settings, we
restore any photograph without prior calibration with a target.

The contributions of this paper are summarized as follows:
\begin{itemize}
    \item We decompose the optical aberration into blur and warp components and in particular, characterize the blur with local 2D Gaussian kernels with seven parameters. We validate this model with the PSFs measurements of \cite{bauer18mtf};
    \item we sequentially compensate the blur and the warp.
    We apply the blind deblurring algorithm of \cite{delbracio21polyblur} to sharpen the image, showcasing its effectiveness for optical aberration correction, and then
    remove the remaining color fringes with a novel 2-channel CNN trained to 
    minimize the image residual between the red/blue and green channels;
    \item quantitative experiments on both synthetic and real images
    show that our method accurately compensates both the blur and the colored
    edges misalignments caused by the PSF. In particular it is 20 times faster and has 100 times less parameters than the current state of the art; and
    \item we show that our blind approach generalizes to real images even competing with commercial image editing software running in a non-blind setting. Our method processes a 12 megapixels image in 1 second on a GPU with a non-optimized code.
\end{itemize}

\section{Related work}

Knowing the PSF associated to an image or a lens may be useful for two tasks: accurately evaluating the lens quality and removing the lens blur with a non-blind deblurring algorithm. The PSF may be estimated from a single photograph of a calibration target or from natural images.
Trimeche \etal\cite{trimeche05multichannel} and Joshi \etal\cite{joshi08sharpedge}
take raw photographs of targets with contrasted edges, \eg a checkerboard,
and solve an optimization problem to predict a grayscale local filter.
The same idea is proposed by Brauers \etal\cite{brauers10direct}, Delbracio 
\etal\cite{delbracio12nonparametric} and Heide \etal\cite{heide13simplelens}
who use carefully designed 
noise patterns to facilitate the optimization
and achieve sub-pixel grayscale filters estimation.
Instead of using edge and noise patterns, Schuler \etal\cite{schuler11nonstationnary} 
and Bauer \etal\cite{bauer18mtf} take photographs of LED panels, 
which allow them to directly observe the local PSFs without 
any optimization, simply by recording how the
white LED dots become colored spots in the images.

All these techniques may predict accurate estimates of the PSF but are only 
valid for specific lens settings and for a sparse set of
locations in the image, making them unsuitable at non-measured pixel 
locations or lens settings.
A few approaches intend to fill this void:
Kee \etal\cite{kee11modeling} and Shih \etal\cite{shih12simulations} 
interpolate the PSF for various focal length/aperture aperture pairs 
by fitting a spatial Gaussian model and
Hirsch and Sch\"olkopf~\cite{hirsch15selfcalibration} predict
RGB filters at unknown locations on the field of view with a kernel method.

However, if the goal is enhancing the image sharpness, blind kernel
estimates designed to achieve the best deblurring, \ie without being  faithful representations of the true local blurs, may suffice.
For instance Joshi \etal\cite{joshi08sharpedge} propose a variant of their target-based
approach by assuming the latent sharp image has ideal step edges.
Schuler \etal\cite{schuler12blind} predict 
a set of RGB linear filters covering the image, hypothesizing symmetries of the PSF,
which is most of the time an inaccurate oversimplification for real lenses \cite{dube17good}, and Yue \etal\cite{yue15geometric}
and Sun \etal\cite{sun17chrosschannel} additionally posit sharpness 
of the green channel, which is also an aggressive approximation when 
looking at real lens measurements~\cite{bauer18mtf}.
Heide \etal\cite{heide13simplelens} adopt instead a prior on the color and
the location of edges across the color channels.
After PSF estimation, correction boils down to non-blind deblurring
by solving an inverse problem \cite{krishnan09fast}, or 
learned with a CNN\cite{li21universal}.
In this paper, we adopt a 2D Gaussian model to approximate the
local blur caused by the PSF, 
which is validated by observations of~\cite{kee11modeling} 
and that can be efficiently estimated from a single image~\cite{delbracio21polyblur}.
%
Furthermore, \cite{delbracio21polyblur} shows that no prior is needed
to achieve satisfactory deblurring results with these simple kernels.

Blur is only one facet of a PSF, which
also warps the color planes of a photograph, resulting in color fringes next to the edges.
Boult and Wolberg~\cite{boult92correction} and Kang \cite{kang07automatic}
align the red and blue channels with the
green one by means of a radial warp model.
Chang \etal\cite{chang13correction} do not suppose any model on the warp and
instead remove the fringes with a linear filter applied in the neighborhood 
if the most salient edges, in the red/green and blur/green image residuals.
These image residuals contain all the information to characterize
these colored artifacts and are used in the present work to train a CNN, 
a non-linear variant of \cite{chang13correction}.

\section{Local PSF parametric model}

\subsection{Optical aberrations model}

In the absence of diffraction, which is a realistic assumption for
usual aperture sizes, typically below $f/11$, the PSF 
is the combination of the optical aberrations. The Seidel 
theory~\cite{tang13what} decomposes them into five monochromatic aberrations:
spherical, coma, astigmatism, field curvature and geometric distortion, and two
chromatic aberrations: lateral and longitudinal.

\begin{figure}[t]
    \centering
    \includegraphics[width=0.9\textwidth]{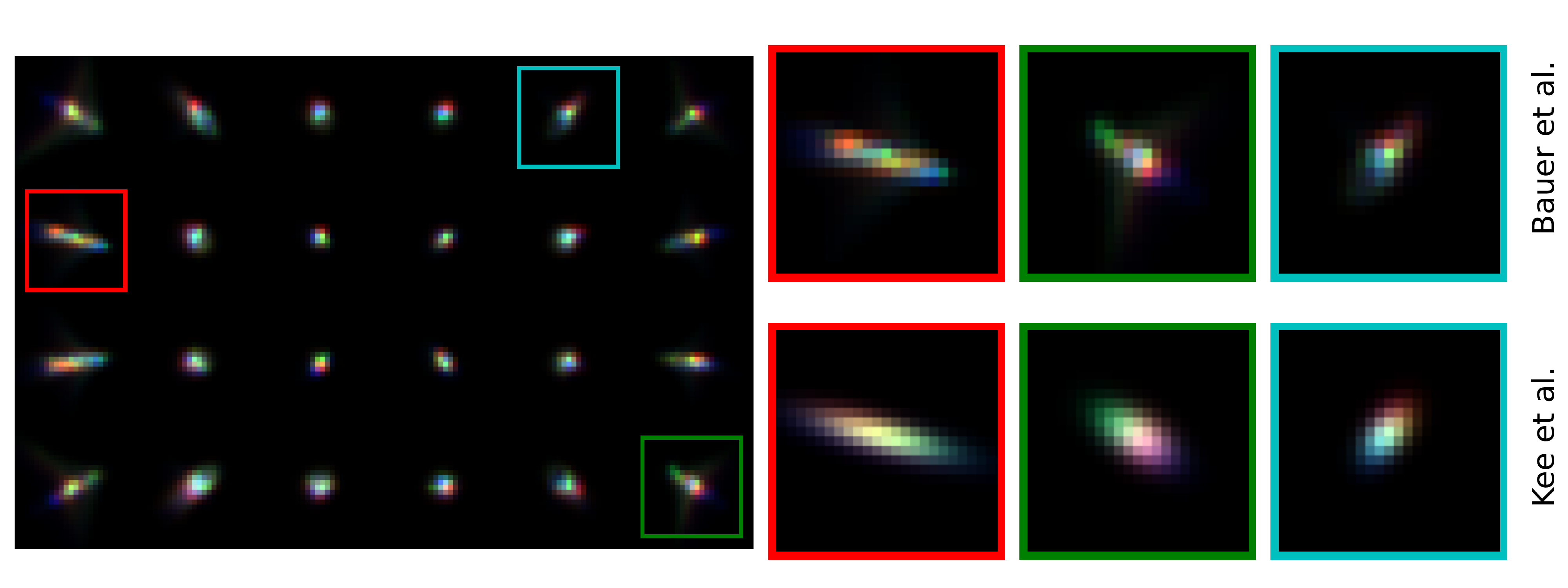}    
    \caption{A $4\times6$ subset of the Canon EF 16-35mm f/2.8L II USM PS lens
    PSF measurement of Bauer \etal\cite{bauer18mtf}
    at maximal aperture and shortest focal length, 
    a panel of three zoomed local kernels
    and the Gaussian approximations of Kee \etal\cite{kee11modeling}. The 
    spots, despite being non-parametric functions of the
    field of view, may be reasonably approximated with spatial Gaussian filters.}
    \label{fig:show_psf}
\end{figure}

The combination of the first four monochromatic and the longitudinal aberrations
boils down to converting a point in the ideally focused image   into
a spot whose size depends on the wavelength and its position on the 
focal plane~\cite{kang07automatic}.
Geometric distortion bends parallel lines and necessitates 
two or more images to calibrate the camera~\cite{zhang00calibration}, and is
thus not addressed in this presentation.
However, lateral aberrations are also  geometric transformations, but which warp differently each color component of an edge, leading to
visible colored fringes~\cite{chang13correction}. 
Figure~\ref{fig:show_psf} illustrates a PSF measurement of a real lens obtained by Bauer~\etal\cite{bauer18mtf}.

Kang\cite{kang07automatic} already proposed a forward model for optical aberrations with simultaneous blur and color warp. Chang~\etal\cite{chang13correction} set an order, that we follow in this paper,
by running a sharpening stage prior to edge correction.
From the above analysis, and considering also degradation caused by the sensor, 
(mosaicking, noise and saturation), we derive the following raw 
image formation model for a single color channel $c=(R,G,B)$:
\begin{equation}\label{eq:rawformationmodel}
    r_c = s \circ m_c \left(g_c \circ w_c(u_c) + \varepsilon\right) 
    \;\;\text{with}\;\;\varepsilon\sim\mathcal{N}(0, \alpha g_c \circ w_c(u_c) + \beta),
\end{equation}
where $u_c$ and $r_c$ are the sharp and raw color planes, $w_c$ is 
the inter-color warp caused by lateral chromatic aberrations (recall that
we neglect geometric distortion in this presentation), 
$g_c$ is the spatially-varying blur caused by the remaining aberrations,
$\circ$ is the composition operator,
$m_c$ is the decimation caused by the mosaicking filter, $s$ is the
sensor saturation and $\varepsilon$ is the image noise modeled
with the heteroscedastic normal model of~\cite{foi08practical}, parameterized
with shot and read noise weights $\alpha$ and $\beta$.
We call $v$ the denoised and demosaicked version of the raw image $r$, 
which should therefore be close to the RGB aberrated image {\em before} mosaicking
and degradation with noise.

\begin{figure}[t]
    \centering
    \begin{tabular}{cc}
        \begin{subfigure}[]{0.48\textwidth}
             \centering
             \includegraphics[width=\textwidth]{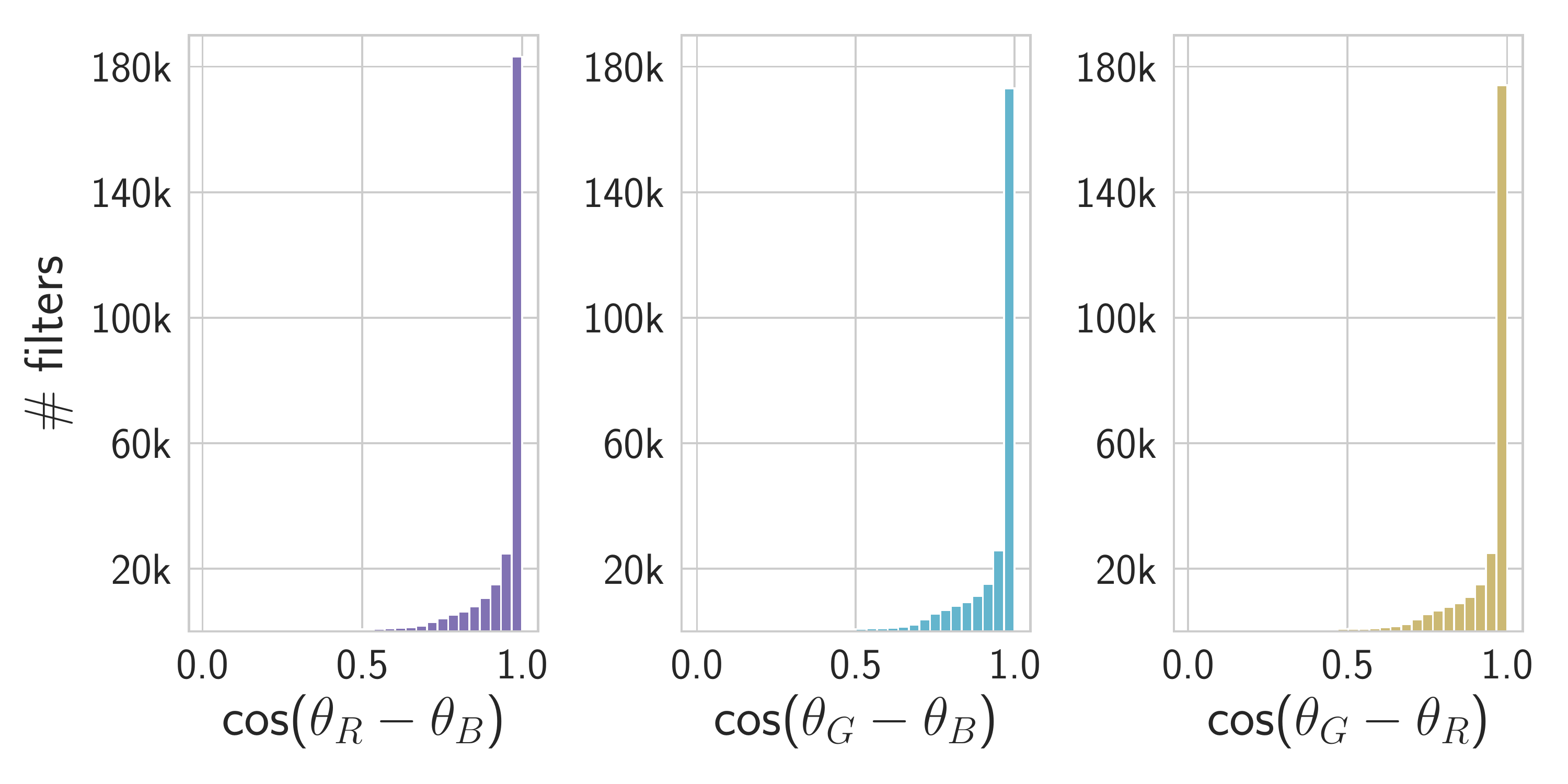}
             \caption{Assumption on $\theta_c$.}
         \end{subfigure} &
         \begin{subfigure}[]{0.48\textwidth}
             \centering
             \includegraphics[width=\textwidth]{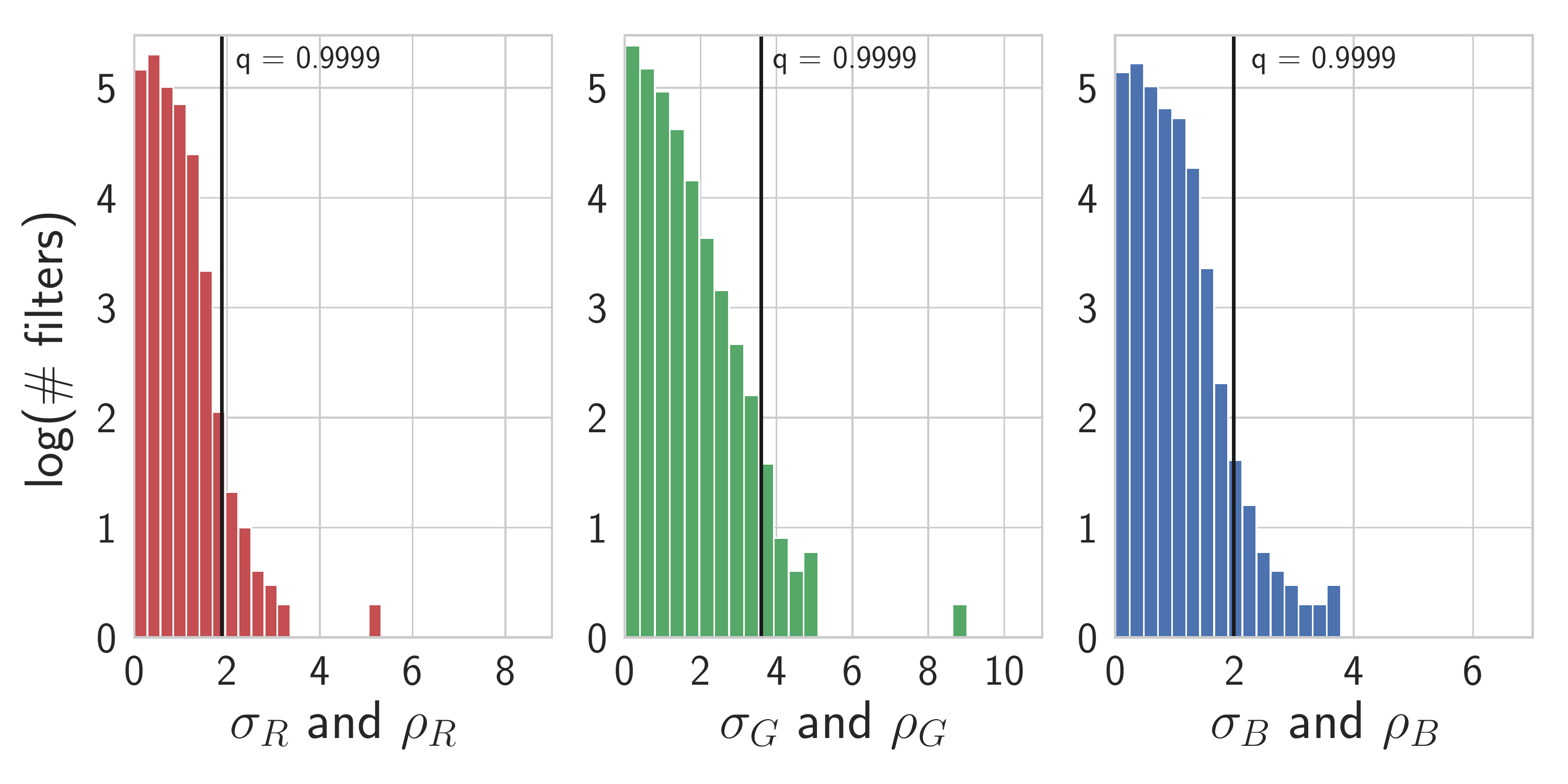}
             \caption{Assumption on $\sigma_c$ and $\rho_c$.}
         \end{subfigure}
    \end{tabular}
    \caption{Experimental validation of the assumptions
    on the parameter triplets $(\theta_c, \sigma_c, \rho_c)$. 
    Left: We measure the similarity $\theta_R$,
    $\theta_G$ and $\theta_B$ and show the strong correlation 
    across the color channel directions. 
    Right: We compute the histogram
    of $\sigma_c$ and $\rho_c$ ($c=R,G,B$) and its 0.9999-th quantile,
    and show that almost all standard deviation
    values are under 4.}
    \label{fig:assumptions}
\end{figure}

\subsection{Blur parametric approximation}

An image of the local blur may be obtained with photographs
of a known reference like a target~\cite{trimeche05multichannel, joshi08sharpedge, delbracio12nonparametric}. It gives an
accurate estimate of the blur but only in a controlled 
environment with special gear.
We instead follow Kee \etal\cite{kee11modeling} and approximate
the local blur $g_c$ in a color channel ($c=R,G,B$) 
with a zero-mean 2D Gaussian filter. Blur estimation thus 
boils down to a direct blind estimation of
few parameters 
from any photograph.

\noindent\textbf{Parametric monochromatic aberrations.}
A zero-mean 2D Gaussian is fully characterized by three parameters: 
the angle of the principal direction $\theta$
and the standard deviation values $\sigma$ in $\theta$ and $\rho$ in the direction
$\theta + \pi/2$:
\begin{equation}\label{eq:2dgaussian}
    k(x) = \left[2\pi\det(\Sigma)\right]^{-\frac{1}{2}}
    \exp{\left(-\frac{1}{2}x^\top\Sigma^{-1}x\right)},
\end{equation}
for the locations $x$ in the support of $k$ and 
with a covariance matrix $\Sigma$:  
\begin{equation}\label{eq:covariance}
    \Sigma = R(\theta)^\top
    \begin{bmatrix}
    \sigma^2 & 0 \\ 0 & \rho^2
    \end{bmatrix}
    R(\theta),
\end{equation}
where $R(\theta)$ is the 2D rotation matrix of angle $\theta$.

\noindent\textbf{Parametric longitudinal chromatic aberrations.}
Incorporating the contribution of longitudinal chromatic aberrations yields a kernel $k_c$, or equivalently a 
triplet $(\theta_c, \sigma_c, \rho_c)$, for each color $c=(R,G,B)$.
Thus, our RGB parametric local blur model has nine parameters.

\noindent\textbf{Bounding the parameters with real data.}
We use the real PSF measurements by Bauer \etal\cite{bauer18mtf}
and the non-blind Gaussian approximation technique 
of Kee \etal\cite{kee11modeling}, both
shown in Figure~\ref{fig:show_psf},
to reduce the set of parameters. Bauer \etal took
photographs with 70 lens settings of a $52\times78$-point LED array,
which yields about 280,000 local RGB PSF measurements $(g_R, g_G, g_B)$.
Since the aperture is kept between $f/1.4$ and $f/5.6$, the
contribution of diffraction to the local spot is negligible. 

Following Kee {\em et al.}, we compute the covariance matrices of the 
kernels $k_c$  that best fit the  measurements $g_c$ $(c=R,G,B)$,
and whose eigendecomposition return
the triplets $(\theta_c, \sigma_c, \rho_c)$.
From this large corpus of triplets, we draw two conclusions: 
(i) the direction $\theta_c$ is roughly the same for each color $c$,
and (ii) the standard deviation values $\sigma_c$ and $\rho_c$
are contained in the segment $[0.2, 4]$.
The first observation limits the actual number of parameters to
be estimated to only seven, the information on the principal
direction being contained in a single scalar $\theta$,
whereas the second observation ensures that we can use the fast blind 
deblurring technique of \cite{delbracio21polyblur} 
to predict $k_c$.

We experimentally validate these claims by first
computing the cosine similarity of the pairs of eigenvectors directed
by $\theta_c$ of the approximate filters $k_c$ ($c=R,G,B$).
We show in Figure~\ref{fig:assumptions}~(a) that these vectors
are always aligned, confirming our first observation.
Second, we plot in Figure~\ref{fig:assumptions}~(b)
the cumulative distribution function
of the standard deviation values $\sigma_c$ and $\rho_c$
and show that
only a negligible amount of candidates are above 4.
We also see that a realistic floor value is at 0.2, 
thus suggesting the standard deviations for modeling
realistic parametric lens blurs are within a segment $[0.2, 4]$,
validating our second claim.
In conclusion, we can reasonably adapt
the blind deblurring technique of Delbracio \etal\cite{delbracio21polyblur}
to estimate a local PSF blur with 
only seven blur parameters.

\section{Proposed method}

We decompose the image into
patches, \eg with 25\% or 50\% overlap, in which we assume that the blur is 
uniform, and we remove the local PSF in two steps.
We first remove the 
local uniform blur with the blind Gaussian deblurring
technique of \cite{delbracio21polyblur}. Second, we eliminate the colored
artifacts caused by the warp next the salient deblurred edges
using a CNN, which is inspired on the method \cite{chang13correction}.
The selected  deblurring  and colored artifact correction methods strike a good compromise
between speed and accuracy. Other combinations of methods were also considered leading to worse results or much slower methods, \eg classical registration techniques for lateral aberration removal~\cite{kang07automatic}, or much slower methods, for instance recent CNNs for deblurring~\cite{li21universal}.

Algorithm~\ref{alg:method} summarizes our approach for restoring a
single patch.
After all the patches are deblurred and 
processed by the CNN, we put them back to their
initial locations in the image using a
Hamming window to limit fusion artifacts.

\begin{algorithm}[t]
\KwData{Aberrated $v$, coefficients $(C,\sigma_b)$, estimator $\phi_\nu$}
\KwResult{Aberration-free $\widehat{u}$}
Compute blur direction $\theta$ from $v_G$ with Eq.~\eqref{eq:thetapolyblur}\;
Compute blur standard deviations $\sigma_c$ and $\rho_c$ $(c=R,G,B)$ with Eq.~\eqref{eq:stdpolyblur}\;
Compute approximate filter $k_c$ $(c=R,G,B)$ with Eqs.~\eqref{eq:2dgaussian} and \eqref{eq:covariance}\;
Compute approximate inverse filter $p(k_c) = -3(k_c \ast k_c) - k_c + 3\delta$ $(c=R,G,B)$\;
Compute deblurred image $z_c$ $(c=R,G,B)$ with Eq.~\eqref{eq:deblurring}\;
Compute aligned channel $\widehat{u}_c$ $(c=R,B)$ with Eq.~\eqref{eq:cnn}\;
Build $\widehat{u} = [\widehat{u}_R, z_G, \widehat{u}_B]$\;
\caption{Proposed PSF removal method}
\label{alg:method}
\end{algorithm}

\subsection{Blind Gaussian deblurring}
\label{subsec:monochromatic}

As explained above, the combination of the
monochromatic and longitudinal aberrations is a 
spatially-varying blur. We split the image into overlapping patches 
where the local blur is supposed uniform, and predict a zero-mean 
Gaussian kernel for which we approximate a deconvolution filter, adapting the procedure of \cite{delbracio21polyblur}.
In brief, this technique quickly estimates the parameters 
$(\rho, \sigma, \theta)$ from a blurry grayscale image, and run 
an approximate inverse filter for the corresponding 2D Gaussian kernel.
It is particularly effective for ``mild'' blurs that may be captured
by Gaussian kernels with standard deviation under 4.

This blind deblurring technique is valid in our context
since PSFs are mostly small blurs according to the previous section
and previous art~\cite{kee11modeling, schuler11nonstationnary, bauer18mtf}. 
The authors of \cite{delbracio21polyblur} thus demonstrate that
their approach achieves similar result to that of CNNs, but for
a fraction of the speed and memory. We show in this work it is well suited
for lens blur removal.
Since, according to
our analysis, the blur orientation is the same for all color channels we find $\theta$ by arbitrarily 
computing the infinite norm of the directional derivative of the green channel and picking the direction with the smallest 
value, \ie where the blur is the strongest:
\begin{equation}\label{eq:thetapolyblur}
    \theta = \argmin_\varphi \|\nabla_\varphi n(v_G)\|_\infty,
\end{equation}
where $\nabla_\varphi v = \cos(\varphi)\nabla_x v + \sin(\varphi)\nabla_y v$, $\nabla_x$ and $\nabla_y$ are the horizontal
and vertical derivative operators, 
and $n$ is a normalization function 
detailed in \cite{delbracio21polyblur} and 
in the supplemental material.
For the range of standard deviation values we are interested with, 
Delbracio \etal\cite{delbracio21polyblur} empirically show that
there exists an affine relationship between the
variance of a Gaussian blur and the infinite norm of the image gradients
in its principal directions $\theta$ and $\theta + \pi/2$. 
Let $C$ be the slope and $\sigma_b$ be 
the intercept of this model. The empirical affine model reads
\begin{equation}\label{eq:stdpolyblur}
    \sigma_c = 
    \sqrt{\frac{C^2}{\ninf{\theta}{n(v_c)}^2} - \sigma_b^2}
    \;\;\text{and}\;\;
    \rho_c =
    \sqrt{\frac{C^2}{\ninf{\theta+\frac{\pi}{2}}{n(v_c)}^2} - \sigma_b^2},
\end{equation}
where $c\in \{R,G,B\}$ and $\theta$ is the direction 
previously computed.
The hyperparameters are tuned with the protocol of \cite{delbracio21polyblur}.
Minimizing with the linear programming algorithm the sum of $\ell_1$ differences between the norm
of the gradient and the variance for 600 synthetic 
blurry images and known corresponding Gaussian filters
yields $C=0.415$ and $\sigma_b=0.358$
for demosaicked images before gamma correction,
and $C=0.371$ and $\sigma_b=0.453$ for JPEG images.

The resulting triplet ($\theta, \sigma_c, \rho_c$) is used to build
the covariance matrix defined in Eq.~\eqref{eq:covariance} and thus the
2D Gaussian kernel $k_c$ ($c=R,G,B$).
As in \cite{delbracio21polyblur}, we carry out non-blind deblurring
by computing the approximate inverse filter
$p(k)=-3(k\ast k)-4k+3\delta$ ($\delta$ denotes the Dirac filter),
and deconvolve each color channel $c$ $(c=R,G,B)$ with:
\begin{equation}\label{eq:deblurring}
    z_c = p(k_c) \ast v_c.
\end{equation}
We have also tried an inverse filter obtained with Fourier transform, \eg\cite{eboli20end2end}, but noticed that the filter $p(k_c)$ ($c=R,G,B$)
achieves better results in our experiments.
Each $h\times w$ image $z_c$ is a sharp version of $v_c$, however due to lateral chromatic aberration,
the red and blue channels still have shifted edges compared to their counterparts in the aberration-free image $u$,  
which results in artifacts in the vicinity of contrasted and sharp edges.

\subsection{Red and blue edge correction} 

Lateral chromatic aberrations introduce a shift between the color channels.
Usual techniques for removing these colored artifacts use parametric red-to-green and blue-to-green warp models,
for instance taking the form of a global radial 
transformation~\cite{boult92correction, kang07automatic} 
or local translations~\cite{schuler12blind, yue15geometric}.
In this context registration is hard since different color 
channels may have different edge profiles and
in these contrasted areas demosaicking may produce 
incorrect color predictions, preventing perfect edge alignment and resulting in residual edge artifacts.
Modeling the warp thus seems to be a harder problem than the original one.
Conversely, we follow Chang \etal\cite{chang13correction} 
remkarking that lateral aberrations result in color fringes next
to the most salient edges; Filtering the edges, without
any explicit model on the warp or information on the edge location, 
is enough for effective correction. In this work we propose a residual CNN,
that takes as input $z_G$ and $z_R$
or $z_B$ and returns an image $\widehat{u}_R$ or $\widehat{u}_B$
whose edges should be aligned with those of $z_G$.
If we call this CNN $\phi$ with parameter $\nu$, our approach reads for $c=R,B$:
\begin{equation}\label{eq:cnn}
    \widehat{u}_c = z_c - \phi_\nu(z_c, z_G).
\end{equation}
We then combine
$\widehat{u}_R$, $z_G$ and $\widehat{u}_B$ into a single 
restored image. The network
$\phi_\nu$ is a UNet with four convolutional layers of respectively
16, 32, 64 and 64 feature maps in the encoder part and a mirrored
structure in the decoder, each followed by batch normalization
and ReLU activation.

\noindent\textbf{Training of $\phi$.}
For estimating the network parameters $\nu$, we use synthetic supervisory data.
We follow Brooks \etal\cite{brooks19unprocessing} to convert $128\times128$ 
JPEG patches into linear RGB ones, just 
after demosaicking, but without 
noise or aberrations. We then apply the forward model \eqref{eq:rawformationmodel} to generate
their aberrated and mosaicked raw counterparts. 
We sample orientations in $[0, \pi)$, and standard deviations in $[0.2, 4]$
to build an RGB Gaussian kernel to blur 
a given ``unprocessed'' training image $u$ from the DIV2K
dataset. Then  translate the red and blue channels with sub-pixel shifts sampled in $[-4,4]^2$
to model the local lateral chromatic aberration, add 
Poissonian-Gaussian noise, mosaick with the Bayer filter
and clip its pixel values between 0 and
1, ultimately resulting in a raw image $r$.
The translation value range is empirically set after having
observed photographs taken with a couple of different lenses.
Nonetheless, this arbitrary value leads to satisfactory restoration
results in real images.
To simulate the modules  preceding the optical aberration brick 
in any image processing pipeline~(see Fig.~\ref{alg:method}), we denoise and demosaick $r$ 
respectively with the bilateral filter~\cite{tomasi1998bilateral} and demosaicnet~\cite{gharbi16deep} to predict an
aberrated RGB image $v$.
We deblur $v$ by 
removing the blur with Eqs.~\eqref{eq:thetapolyblur} to
\eqref{eq:deblurring} to predict a sharp version $z$ 
with aberrated edges.

As demonstrated by Chang \etal\cite{chang13correction}, 
the chroma images $z_R-z_G$ and 
$z_B-z_G$ isolate the lateral chromatic
aberrations and are sufficient to remove the colored artifacts.
Thus, instead of training our model to minimize a loss 
of the sort $\| \widehat{u} -u \|_1$ as usual, 
we force $\phi$ to minimize these quantities 
for $N$ synthetic image pairs $(u^{(i)}, v^{(i)})$ with the training
loss
\begin{equation}\label{eq:training}
    \sum_{i=1}^N \sum_{c\in\{R,B\}}
    \left\| \left(u^{(i)}_c - u^{(i)}_G\right) - \left(z^{(i)}_c - \phi_\nu(z^{(i)}_c, z_G^{(i)}) - z_G^{(i)}\right)\right\|_1,
\end{equation}
where $z^{(i)}_c = p(k_c) \ast v^{(i)}_c$ ($c=R,G,B$).
Since the roles of the red and blue channels are symmetric,
we have $2N$ supervisions from $N$ pairs $(u^{(i)}, v^{(i)})$ ($i=1,\dots,N$).
We minimize Eq.~\eqref{eq:training} with the Adam optimizer whose
initial learning rate is set to $3\times10^{-4}$ and is multiplied
by 0.5 when the validation loss plateaus for 10 epochs and with
batch size set to 40.

\section{Experiments}

\subsection{Blind grayscale PSF removal}

\begin{figure}[t]
    \centering
     \begin{tabular}{ccccc}
        \begin{subfigure}[b]{0.19\textwidth}
         \centering
         \includegraphics[width=\textwidth]{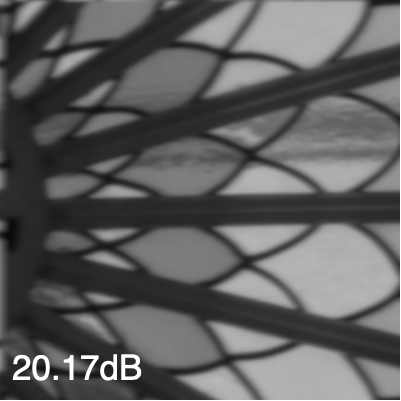}
         \caption{Blurry.}
     \end{subfigure}  &  
        \begin{subfigure}[b]{0.19\textwidth}
         \centering
         \includegraphics[width=\textwidth]{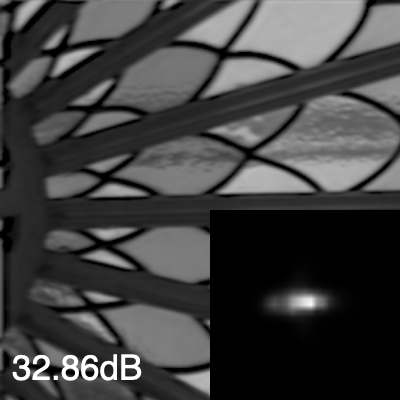}
         \caption{GT's kernel.}
     \end{subfigure}  &  
    \begin{subfigure}[b]{0.19\textwidth}
         \centering
         \includegraphics[width=\textwidth]{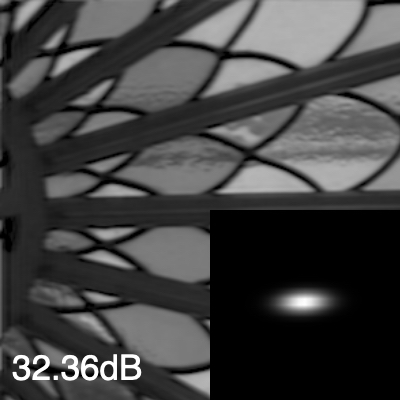}
         \caption{\cite{kee11modeling}'s kernel.}
     \end{subfigure}  & 
            \begin{subfigure}[b]{0.19\textwidth}
         \centering
         \includegraphics[width=\textwidth]{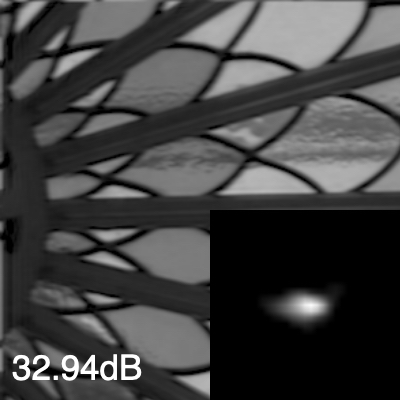}
         \caption{\cite{anger19blind}'s kernel.}
     \end{subfigure}  &  
             \begin{subfigure}[b]{0.19\textwidth}
         \centering
         \includegraphics[width=\textwidth]{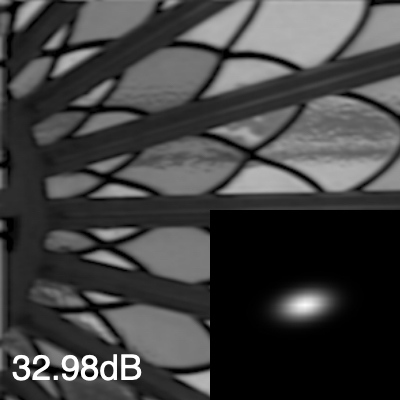}
         \caption{Our's kernel.}
     \end{subfigure}
    \end{tabular}
    \caption{
    Qualitative result for blind deblurring with different kernel estimators. 
    From left to right: The synthetic blurry image, the version deblurred with 
    the ground truth kernel from \cite{bauer18mtf}, the oracle Gaussian 
    approximation \cite{kee11modeling}, the parametric kernel from 
    \cite{anger19blind} and our estimate. We use the polynomial $p$ previously defined
    to achieve non-blind deblurring. All the techniques, except that of 
    Kee~\etal~\cite{kee11modeling} achieve similar results but ours is blind and fast.
    }
    \label{fig:deblurring qualitative}
\end{figure}

We first measure the ability of the parametric estimation technique
to help deblurring a real-world non-parametric PSF for a single color channel
(the impact of lateral chromatic aberrations is kept for later in 
this presentation). 
We compute blur estimates with a panel of blur estimation techniques
including ours, and quantitatively evaluate their impact on deblurring.

\begin{figure}[t]
    \centering
     \begin{tabular}{cc}
        \begin{subfigure}[b]{0.48\textwidth}
         \centering
         \includegraphics[width=\textwidth]{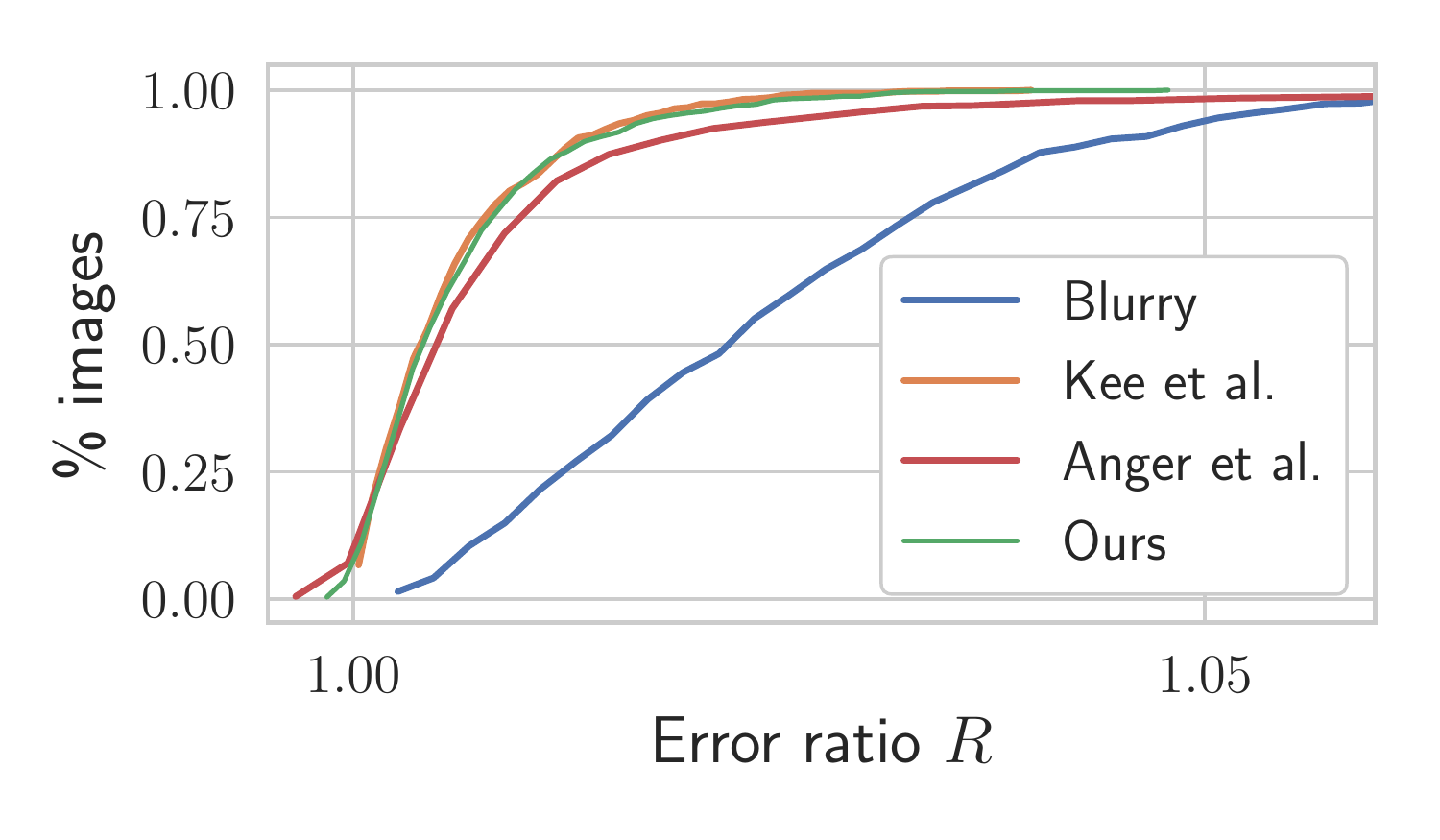}
         \caption{Blind deblurring.}
     \end{subfigure}  &  
    \begin{subfigure}[b]{0.48\textwidth}
         \centering
         \includegraphics[width=\textwidth]{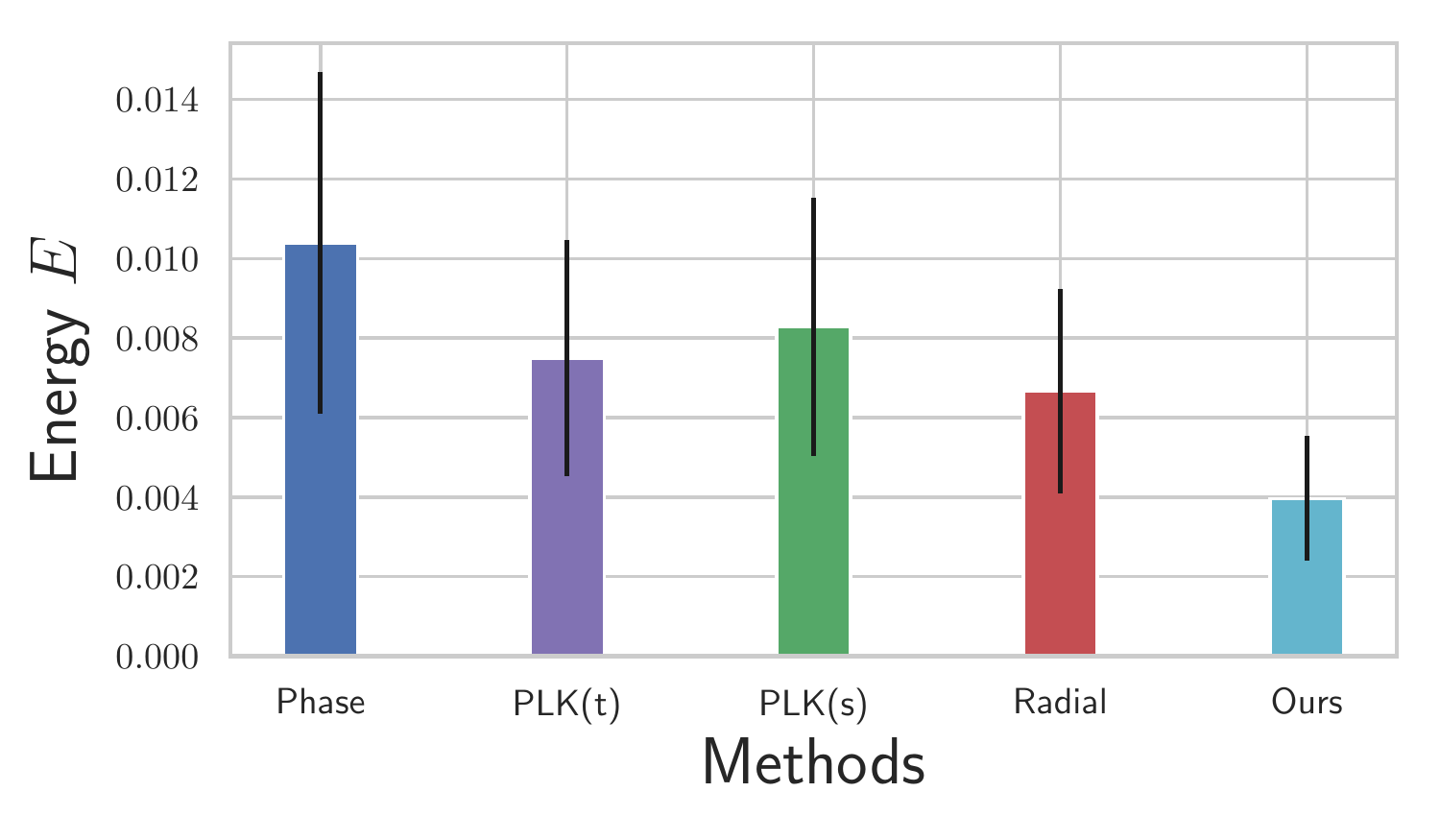}
         \caption{Edge correction.}
     \end{subfigure}
    \end{tabular}
    \caption{
    Quantitative analysis of the blind deblurring and the 
    edge corrections modules with the metrics $R$ and $E$
    of Eqs~\eqref{eq:ratio} and~\eqref{eq:energyE}.
    Left: Comparison of the SSIM ratios $R$ in Eq.~\eqref{eq:ratio}
    for kernels estimated as by Kee \etal\cite{kee11modeling}, Anger
    \etal\cite{anger19blind} and with our approach 
    (the more on the left, the better).
    Our blind method competes with the non-blind technique
    of Kee {\em et al}.
    Right: Comparison of the energy $E$ in Eq.~\eqref{eq:energyE} from Heide \etal\cite{heide13simplelens} for edge corrections estimated by phase correlation \cite{leprince07automatic}, the pyramid Lucas-Kanade (PLK) algorithm of~\cite{baker04lucas} predicting translations  and similarities (PLK(t) and PLK(s)), the radial model of~\cite{kang07automatic}
    and our CNN.
    Our approach achieves the best quantitative result.
    }
    \label{fig:graypsfestimation}
\end{figure}

We convolve grayscale images $u$ with the green components $g_G$
of the local PSFs of  Bauer 
\etal\cite{bauer18mtf} to  generate 
blurry images $v$, from which we predict a blur kernel $\widehat{g_G}$ with 
various kernel estimation techniques. We then compute a
deconvolution filter $p(\widehat{g_G})$ and estimate a deblurred version $p(g_G) \ast v$
for each kernel estimation method in our panel composed 
of the non-blind parametric model of Kee \etal\cite{kee11modeling}
and the blind non-parametric algorithm of Anger \etal\cite{anger19blind}.
We quantitatively compare the performance of the blur estimators with the
SSIM ratio of Kee \etal comparing the relative quality of the image
deblurred with the ground-truth kernel $g_G$ 
over that restored with $\widehat{g_G}$:
\begin{equation}\label{eq:ratio}
    R(\widehat{g_G},g_G) = \frac{\text{SSIM}[p(g_G) \ast v, u] + 2}{\text{SSIM}[p(\widehat{g_G}) \ast v, u] + 2}.
\end{equation}
Since the kernels of Bauer \etal may not be centered in zero, we adopt the 
ground-truth shifting strategy of Levin \etal\cite{levin09understanding}
and crop the 15 pixel on the borders to compute $\text{SSIM}[p(g_G) \ast v, u]$.
Figure \ref{fig:graypsfestimation}~(a) shows the plots of the ratios $R$
for the different kernel estimators on 870 synthetic images of 
size $400\times400$.
The non-blind parametric technique of Kee \etal is an upper-bound to
ours and logically achieves the best result,
nonetheless we are just under it with a marginal gap, and in a blind
fashion. We also
exceed the performance of the non-parametric algorithm
of Anger {\em et al.}, validating our blind Gaussian model for PSF removal. Figure~\ref{fig:deblurring qualitative} shows
a deblurring example for different kernel estimates.

\subsection{Lateral chromatic aberration compensation}
We now validate the CNN $\phi$ to
correct the lateral chromatic aberrations.
However, to our knowledge, there is no benchmark or quantitative metric for 
this specific task. As a result, we
have found that computing the norm of the image prior of Heide
\etal\cite{heide13simplelens} favoring aberration-free solutions, 
was the most relevant existing metric for this evaluation.
Given an image $z$,
we predict the red and blur corrected planes $\widehat{u}_R$
and $\widehat{u}_B$, compute their horizontal and vertical gradients
with $\nabla_x$ and $\nabla_y$, and evaluate the following energy:
\begin{equation}\label{eq:energyE}
    E(\widehat{u}_R, \widehat{u}_B, z_G) = \sum_{c=R,B}\sum_{j=x,y}\|(\nabla_j z_G)/z_G - (\nabla_j\widehat{u}_c)/\widehat{u}_c\|_1,
\end{equation}
where the division is pixelwise. It may be seen as normalized variants of the 
color residuals of Chang \etal\cite{chang13correction}.
Note that this quantitative score does not necessitate
a clean ground-truth, and thus can be used on real images.
We thus take ten 24 megapixels photographs,
of various environments (shown in the supplemental material), that are denoised and demosaicked with
DxO PhotoLab 5, deblurred with our blind technique, and decomposed
into 400$\times400$ non-overlapping patches, resulting in 1,500 test images.

Figure~\ref{fig:graypsfestimation}~(b) compares 
the performance of our method with a classical radial model~\cite{kang07automatic}, 
and local parametric warps modeled with translations 
predicted with the phase correlation~\cite{leprince07automatic} or the pyramid Lucas-Kanade (PLK)~\cite{baker04lucas} algorithms,
or similarities also predicted with PLK.
Our model achieves the best performance of the panel since
it is trained to compensate the colored residuals.
Note that phase correlation performs the worst among the considered methods, probably because the real blurs can affect differently the phase of different bands. The under-constrained PLK (similarity) method produces slightly worse results than the radial and PLK (translation) methods.
A visual inspection of the restored images (reported in the supplementary material) confirms this quantitative analysis.

\subsection{Real-world examples}

We test our method on real raw images and some datasets for existing
images comparing our results
with those of DxO PhotoLab 5. Figure \ref{fig:teaser} shows a
real 24 megapixels photograph taken with a Sony $\alpha$6000 camera 
and a Sony
FE 35mm $f/1.8$ lens set at maximal aperture to maximize the 
chromatic aberration.
The raw image is denoised and demosaicked with DxO PhotoLab prior
to optical aberration compensation.
We show in the supplemental material additional 
qualitative results for different lenses.

\begin{figure}[t]
    \centering
    \includegraphics[width=\textwidth]{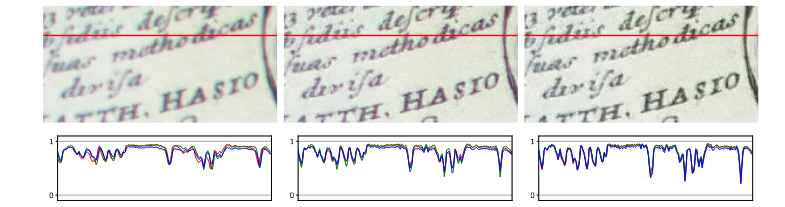}
    \caption{Comparison of lateral chromatic aberration removal from a real raw image. From left to right: The blurry image, the 
    version predicted by $\phi$ trained with the typical loss
    $\| \widehat{u} - u \|_1$ and the estimate from the one trained
    with the loss $\| (\widehat{u}- \widehat{u}_G) - (u - u_G) \|_1$.
    The model trained with the usual regression loss retains
    purplish edges whereas the variant
    gets rid of most of the aberrations.}
    \label{fig:lossimpact}
\end{figure}

\noindent\textbf{Computational efficiency.}
We evaluate the speed of the state-of-the-art CNN from \cite{li21universal}
and our technique to process a 24 megapixel ($6000\times4000$) 
photograph on a NVIDIA 3090 GPU. Our technique takes in average 1.7 seconds
whereas that of \cite{li21universal} takes about 30 seconds on 
the same device. This is explained
by the fact that our network only has 160K parameters for 33.1 gigaflops,
whereas its counterpart counts 17 million parameters for 27.3 teraflops.

\noindent\textbf{Impact of the training loss.}
We train $\phi_\nu$ with a loss minimizing the red-green
and blue-green residuals in the target $u$ and prediction 
$\widehat{u}$ of the form $\| (\widehat{u} - \widehat{u}_G) - (u - u_G) \|_1$, 
which differs from the typical regression loss
$\| \widehat{u} - u\|_1$.
We show in Figure~\ref{fig:lossimpact} the advantage
of the loss~\eqref{eq:training} leveraging the observations
of Chang \etal on chromatic aberrations.
The model trained with the typical regression loss leads 
to purplish edges next to the contrasted edges, \ie the edges across
the three color channels have been aligned but the 
intensities of the red and blue ones do not match that of the green channel,
whereas the one trained with
Eq.~\eqref{eq:training} predicts an image without any color artifact.

\begin{figure}[t]
    \centering
    \begin{tabular}{cccc}
    \begin{subfigure}[b]{0.24\textwidth}
        \centering
        \includegraphics[width=\textwidth]{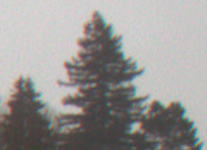}
    \end{subfigure}  &  
    \begin{subfigure}[b]{0.24\textwidth}
        \centering
        \includegraphics[width=\textwidth]{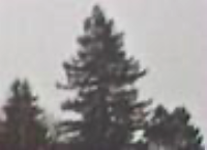}
    \end{subfigure}  &  
    \begin{subfigure}[b]{0.24\textwidth}
        \centering
        \includegraphics[width=\textwidth]{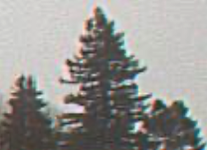}
    \end{subfigure}  &  
    \begin{subfigure}[b]{0.24\textwidth}
        \centering
        \includegraphics[width=\textwidth]{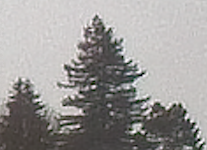}
    \end{subfigure}  \\
    \begin{subfigure}[b]{0.24\textwidth}
        \centering
        \includegraphics[width=\textwidth]{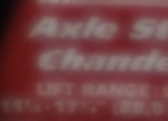}
        \caption{Blurry.}
    \end{subfigure}  &  
    \begin{subfigure}[b]{0.24\textwidth}
        \centering
        \includegraphics[width=\textwidth]{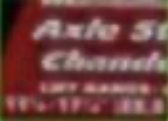}
        \caption{Yue~\etal\cite{yue15geometric}.}
    \end{subfigure}  &  
    \begin{subfigure}[b]{0.24\textwidth}
        \centering
        \includegraphics[width=\textwidth]{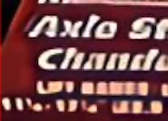}
        \caption{Li~\etal\cite{li21universal}.}
    \end{subfigure}  &  
    \begin{subfigure}[b]{0.24\textwidth}
        \centering
        \includegraphics[width=\textwidth]{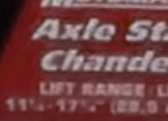}
        \caption{Ours.}
    \end{subfigure} 
    \end{tabular}
    \caption{Comparison of aberration removal from a real JPEG images
    from \cite{kee11modeling} and \cite{heide13simplelens}.
    We eliminate the
    aberrations whereas the competitors retain
    colored edges and cannot restore finer details.}
    \label{fig:jpeg}
\end{figure}

\noindent\textbf{Restoring JPEG images.}
We have assumed so far that the raw image is available.
However, we show that our blind method may also be applied to
JPEG images when only this one is available.
Figure~\ref{fig:jpeg} shows a restoration example from
two images of \cite{kee11modeling} and \cite{heide13simplelens} 
with the techniques
of \cite{yue15geometric, li21universal} 
and ours with the blur estimation
coefficients $(C,\sigma_b)$ calibrated for JPEG images (see Section
\ref{subsec:monochromatic}).
Our method, despite being blind, achieves the best visual
result, predicting correct colors and compensating
the colored edges. 
Since the CNN is trained on linear images, prior to 
restoration we apply an inverse 2.2 gamma curve.

\noindent\textbf{Limitation of the Gaussian model.}
We showed good performance for eight mid-level camera/lens pairs in our experiments.
This guarantees generalization of the Gaussian blur model to that category of photography gear, as claimed by previous art~\cite{kee11modeling, schuler11nonstationnary, bauer18mtf}.
Yet, this model may be
too restrictive in practice, especially for the first-entry lenses for 
which the lens blur may not be captured by
a Gaussian kernel. We show failing examples in the supplementary
material.

\section{Conclusion}

We have proposed a two-stage 
blind method for removing the lens blur,
\ie its PSF, from a JPEG or raw image.
The first module is a blind deblurring technique based on
fast 2D Gaussian filter estimation on overlapping patches.
We have shown that simple parametric kernels are good approximations
of the combination of the monochromatic and longitudinal 
chromatic aberrations.
The second module aligns the red and blue salient edges with
the green ones and thus corrects the lateral chromatic 
aberration.
Experiments have shown that the method generalizes to
real-world images, even in the presence of the
challenging purple fringes.
Our approach is also fast, 
processing a 12 megapixels image
in less than 1 second on a GPU, 
making it suitable for embedding in an ISP pipeline.

\section*{Acknowledgements}

This work was partly financed 
by the DGA Astrid Maturation project ``SURECAVI'' no ANR-21-ASM3-0002,
Office of Naval research grant N00014-17-1-2552. This work was 
performed using HPC resources from GENCI–IDRIS (grant 2022-AD011012453R1).

\bibliographystyle{splncs04}
\bibliography{bibliography}

\begin{thebibliography}{10}
\providecommand{\url}[1]{\texttt{#1}}
\providecommand{\urlprefix}{URL }
\providecommand{\doi}[1]{https://doi.org/#1}

\bibitem{anger19blind}
Anger, J., Facciolo, G., Delbracio, M.: Blind image deblurring using the
  $\ell_0$ gradient prior. Image Processing Online ({IPOL})  \textbf{9},
  124--142 (2019)

\bibitem{baker04lucas}
Baker, S., Matthews, I.A.: Lucas-{K}anade 20 years on: {A} unifying framework.
  International Journal on Computer Vision {(IJCV)}  \textbf{56}(3),  221--255
  (2004)

\bibitem{bauer18mtf}
Bauer, M., Volchkov, V., Hirsch, M., Sch{\"{o}}lkopf, B.: Automatic estimation
  of modulation transfer functions. In: Proceedings of the International
  Conference on Computational Photography {(ICCP)}. pp. 1--12 (2018)

\bibitem{boult92correction}
Boult, T.E., Wolberg, G.: Correcting chromatic aberrations using image warping.
  In: Proceedings of the conference on Computer Vision and Pattern Recognition
  {(CVPR)}. pp. 684--687. {IEEE} (1992)

\bibitem{brauers10direct}
Brauers, J., Seiler, C., Aach, T.: Direct {PSF} estimation using a random noise
  target. In: Digital Photography. {SPIE} Proceedings, vol.~7537, p. 75370.
  {SPIE} (2010)

\bibitem{brooks19unprocessing}
Brooks, T., Mildenhall, B., Xue, T., Chen, J., Sharlet, D., Barron, J.T.:
  Unprocessing images for learned raw denoising. In: Proceedings of the
  conference on Computer Vision and Pattern Recognition {(CVPR)}. pp.
  11036--11045 (2019)

\bibitem{chang13correction}
Chang, J., Kang, H., Kang, M.G.: Correction of axial and lateral chromatic
  aberration with false color filtering. {IEEE} Transactions on Image
  Processing ({TIP})  \textbf{22}(3),  1186--1198 (2013)

\bibitem{delbracio21polyblur}
Delbracio, M., Garcia{-}Dorado, I., Choi, S., Kelly, D., Milanfar, P.:
  Polyblur: {R}emoving mild blur by polynomial reblurring. {IEEE} Transactions
  on Computational Imaging ({TCI})  \textbf{7},  837--848 (2021)

\bibitem{delbracio12nonparametric}
Delbracio, M., Mus{\'{e}}, P., Almansa, A., Morel, J.: The non-parametric
  sub-pixel local point spread function estimation is a well posed problem.
  International Journal on Computer Vision ({IJCV})  \textbf{96}(2),  175--194
  (2012)

\bibitem{dube17good}
Dube, B., Cicala, R., Closz, A., Rolland, J.: How good is your lens?
  {A}ssessing performance with {MTF} full-field displays. Applied Optics
  \textbf{56}(20),  5661--5667 (2017)

\bibitem{eboli20end2end}
Eboli, T., Sun, J., Ponce, J.: End-to-end interpretable learning of non-blind
  image deblurring. In: Proceedings of the European Conference on Computer
  Vision ({ECCV}). pp. 314--331 (2020)

\bibitem{foi08practical}
Foi, A., Trimeche, M., Katkovnik, V., Egiazarian, K.O.: Practical
  {P}oissonian-{G}aussian noise modeling and fitting for single-image raw-data.
  {IEEE} Transactions on Image Processing ({TIP})  \textbf{17}(10),  1737--1754
  (2008)

\bibitem{gharbi16deep}
Gharbi, M., Chaurasia, G., Paris, S., Durand, F.: Deep joint demosaicking and
  denoising. {ACM} Transactions on Graphics (ToG)  \textbf{35}(6),
  191:1--191:12 (2016)

\bibitem{hamilton97interpolation}
Hamilton, J.F., {Adams Jr}, J.E.: Adaptive color plan interpolation in single
  sensor color electronic camera (1997), {US} patent 5629734

\bibitem{heide13simplelens}
Heide, F., Rouf, M., Hullin, M.B., Labitzke, B., Heidrich, W., Kolb, A.:
  High-quality computational imaging through simple lenses. {ACM} Transactions
  on Graphics ({ToG})  \textbf{32}(5),  149:1--149:14 (2013)

\bibitem{hirsch15selfcalibration}
Hirsch, M., Sch{\"{o}}lkopf, B.: Self-calibration of optical lenses. In:
  Proceedings of the International Conference on Computer Vision {(ICCV)}. pp.
  612--620 (2015)

\bibitem{joshi08sharpedge}
Joshi, N., Szeliski, R., Kriegman, D.J.: {PSF} estimation using sharp edge
  prediction. In: Proceedings of the conference on Computer Vision and Patter
  Recognition {(CVPR)} (2008)

\bibitem{kang07automatic}
Kang, S.B.: Automatic removal of chromatic aberration from a single image. In:
  Proceedings of the conference on Computer Vision and Pattern Recognition
  ({CVPR}) (2007)

\bibitem{kee11modeling}
Kee, E., Paris, S., Chen, S., Wang, J.: Modeling and removing spatially-varying
  optical blur. In: Proceedings of the International Conference on
  Computational Photography ({ICCP}). pp.~1--8. {IEEE} Computer Society (2011)

\bibitem{krishnan09fast}
Krishnan, D., Fergus, R.: Fast image deconvolution using hyper-{L}aplacian
  priors. In: Advances in Neural Information Processing Systems ({NeurIPS}).
  pp. 1033--1041 (2009)

\bibitem{leprince07automatic}
Leprince, S., Barbot, S., Ayoub, F., Avouac, J.: Automatic and precise
  orthorectification, coregistration, and subpixel correlation of satellite
  images, application to ground deformation measurements. {IEEE} Transactions
  on Geoscience and Remote Sensing  \textbf{45}(6),  1529--1558 (2007)

\bibitem{levin09understanding}
Levin, A., Weiss, Y., Durand, F., Freeman, W.T.: Understanding and evaluating
  blind deconvolution algorithms. In: Proceedings of the conference on Computer
  Vision and Pattern Recognition ({CVPR}). pp. 1964--1971. {IEEE} Computer
  Society (2009)

\bibitem{li21universal}
Li, X., Suo, J., Zhang, W., Yuan, X., Dai, Q.: Universal and flexible optical
  aberration correction using deep-prior based deconvolution. In: Proceedings
  of the International Conference on Computer Vision {(ICCV)}. pp. 2593--2601
  (2021)

\bibitem{schuler11nonstationnary}
Schuler, C.J., Hirsch, M., Harmeling, S., Sch{\"{o}}lkopf, B.: Non-stationary
  correction of optical aberrations. In: Proceedings of the International
  Conference on Computer Vision ({ICCV}). pp. 659--666 (2011)

\bibitem{schuler12blind}
Schuler, C.J., Hirsch, M., Harmeling, S., Sch{\"{o}}lkopf, B.: Blind correction
  of optical aberrations. In: Proceedings of the European Conference on
  Computer Vision ({ECCV}). pp. 187--200 (2012)

\bibitem{shih12simulations}
Shih, Y., Guenter, B., Joshi, N.: Image enhancement using calibrated lens
  simulations. In: Proceedings of the European Conference on Computer Vision
  ({ECCV}). pp. 42--56 (2012)

\bibitem{sun17chrosschannel}
Sun, T., Peng, Y., Heidrich, W.: Revisiting cross-channel information transfer
  for chromatic aberration correction. In: Proceedings of the International
  Conference on Computer Vision {(ICCV)}. pp. 3268--3276 (2017)

\bibitem{tang13what}
Tang, H., Kutulakos, K.N.: What does an aberrated photo tell us about the lens
  and the scene? In: Proceedings of the International Conference on
  Computational Photography {(ICCP)}. pp. 1--10 (2013)

\bibitem{tomasi1998bilateral}
Tomasi, C., Manduchi, R.: Bilateral filtering for gray and color images. In:
  Proceedings of the International Conference on Computer Vision ({ICCV}). pp.
  839--846 (1998)

\bibitem{trimeche05multichannel}
Trimeche, M., Paliy, D., Vehvilainen, M., Katkovnik, V.: Multichannel image
  deblurring of raw color components. In: Computational Imaging. vol.~5674, pp.
  169--178. SPIE (2005)

\bibitem{wang20practical}
Wang, Y., Huang, H., Xu, Q., Liu, J., Liu, Y., Wang, J.: Practical deep raw
  image denoising on mobile devices. In: Proceedings of the European Conference
  on Computer Vision ({ECCV}). pp. 1--16 (2020)

\bibitem{yue15geometric}
Yue, T., Suo, J., Wang, J., Cao, X., Dai, Q.: Blind optical aberration
  correction by exploring geometric and visual priors. In: Proceedings of the
  conference on Computer Vision and Pattern Recognition {(CVPR)}. pp.
  1684--1692 (2015)

\bibitem{zhang00calibration}
Zhang, Z.: A flexible new technique for camera calibration. {IEEE} Transactions
  on Pattern Analysis and Machine Intelligence ({TPAMI})  \textbf{22}(11),
  1330--1334 (2000)

\end{thebibliography}

\newpage
\renewcommand\thesection{\Alph{section}}
\setcounter{section}{0}

\chapter*{Fast Two-step Blind Optical 
Aberration Correction - Supplementary material}

We provide additional results and companion analyses to those
of the main paper. Section~\ref{sec:deblurring} provides
more details on the blind deblurring algorithm, Section~\ref{sec:cnn} focuses on the proposed CNN and 
its impact on the restoration pipeline, Section~\ref{sec:pipeline} discusses general implementation details of the method, and Section~\ref{sec:images} shows additional qualitative results
for raw images we have taken as well as the JPEG images of
\cite{schuler12blind}.

\section{Deblurring implementation details}
\label{sec:deblurring}

\subsection{Normalizing function}

The normalizing function of Delbracio \etal\cite{delbracio21polyblur} ensures
that the images all have ideal latent edges between 0 and 1.
We have observed that this is a critical component to make the
blur estimation algorithm work.
We use the same function as Delbracio {\em et al.}, defined
by
\begin{equation}
    n(v_G) = \min\left(\max\left(\frac{v_G - v_G[q]}{v_G[1-q] - v_G[q]}, 0\right), 1\right),
\end{equation}
where $v_G[q]$ is the $q$-th quantile of the pixel values in
$v_G$. The quantile value $q$ is set to $0.001$ in all the experiments of this
presentation.

\subsection{Interpolating the angles}

We cannot compute the directional image derivative $\nabla_\varphi n(v_G)$ in all the possible angular directions;
It would be too slow. We follow \cite{delbracio21polyblur} and
actually compute the derivatives for $\varphi$ in $\{0, 30, 60, 90, 120, 150, 180\}^\circ$. The then compute the 
corresponding gradient magnitudes infinite norms $$\|\nabla_\varphi n(v_G)\|_\infty = \max_x|\nabla_\varphi n(v_G)(x)|,$$
and linearly interpolate these values at every $6^\circ$ angle, \ie we predict the infinite norm values for $\varphi$ in
$\{0, 6, 12, \dots, 174, 180\}^\circ$, before
computing the argmax with respect to $\varphi$. We have found that
in practice this strategy was fast and accurate enough to
approximate the real lens blurs.

\subsection{Bounding the standard deviation predictions}

We predict the parameters of the Gaussian approximation of the blur kernel $\sigma_c$ (resp. $\rho_c$) with
\begin{equation}
    \sigma_c = \sqrt{\frac{C^2}{\|\nabla_\varphi n(v_c)\|_\infty^2} - \sigma_b^2},
\end{equation}
as in \cite{delbracio21polyblur}. We however have remarked
that when the magnitude of the gradients was to small, \eg
in textured areas like a tree seen from afar, this equation
was predicting a very large blur, even with the normalization
function.
As discussed in Section \ref{sec:pipeline}, increasing
the patch size may help. To limit this problem Delbracio \etal proposed a clipping strategy. However, in this work we use the following conservative strategy
\begin{equation}
    \sigma_c^\star = \left\{
    \begin{array}{ll}
        0.2 & \mbox{if } \sigma_c > 4 \mbox{ or } V(n(v_c)) < \tau, \\
        \sigma_c & \mbox{otherwise,}
    \end{array}
    \right.
\end{equation}
where $V$ is the variance operator and the  threshold $\tau$ is set to 0.09. This strategy leads to a filter similar to a Dirac impulse, preventing deblurring artifacts in case of ill-blur prediction or ``flat'' patch, \eg a patch with only the sky.
The same technique
is also applied to $\rho_c$.

\section{CNN details}
\label{sec:cnn}

\subsection{Motivation for the loss design}

\begin{figure}[t]
     \centering
     \begin{tabular}{cccc}
    \begin{subfigure}[b]{0.24\textwidth}
        \centering
        \includegraphics[width=\textwidth]{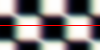}
    \end{subfigure} &  
    \begin{subfigure}[b]{0.24\textwidth}
        \centering
        \includegraphics[width=\textwidth]{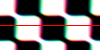}
    \end{subfigure} &
    \begin{subfigure}[b]{0.24\textwidth}
        \centering
        \includegraphics[width=\textwidth]{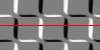}
    \end{subfigure} &
    \begin{subfigure}[b]{0.24\textwidth}
        \centering
        \includegraphics[width=\textwidth]{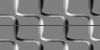}
    \end{subfigure} \\
    \begin{subfigure}[b]{0.24\textwidth}
        \centering
        \includegraphics[width=\textwidth]{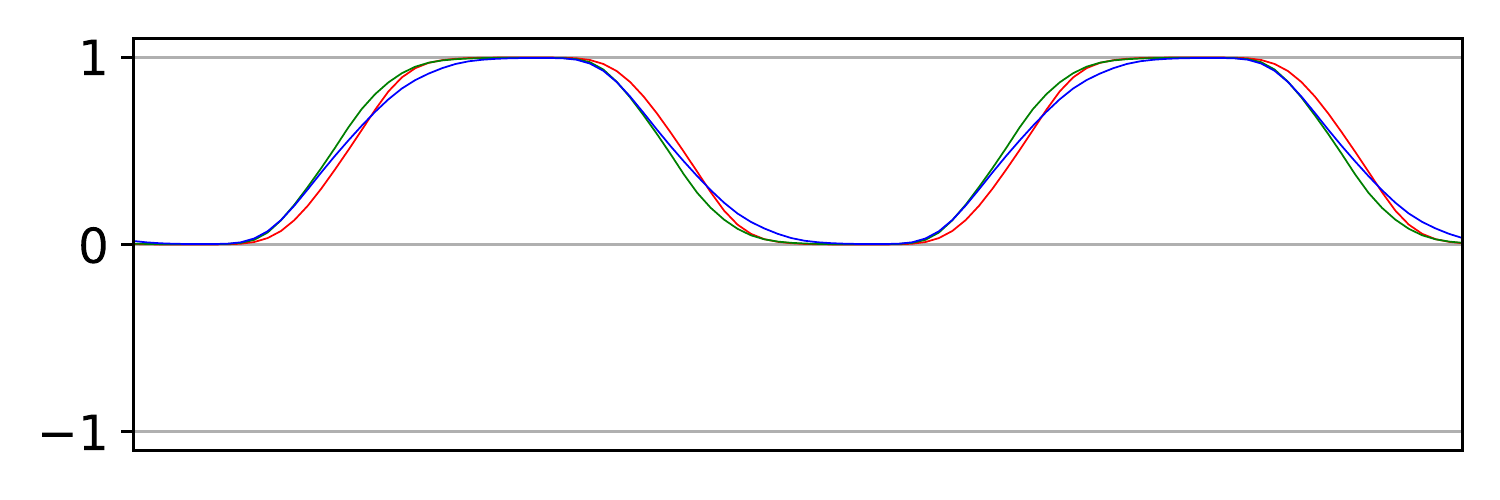}
    \end{subfigure} &
    \begin{subfigure}[b]{0.24\textwidth}
        \centering
        \includegraphics[width=\textwidth]{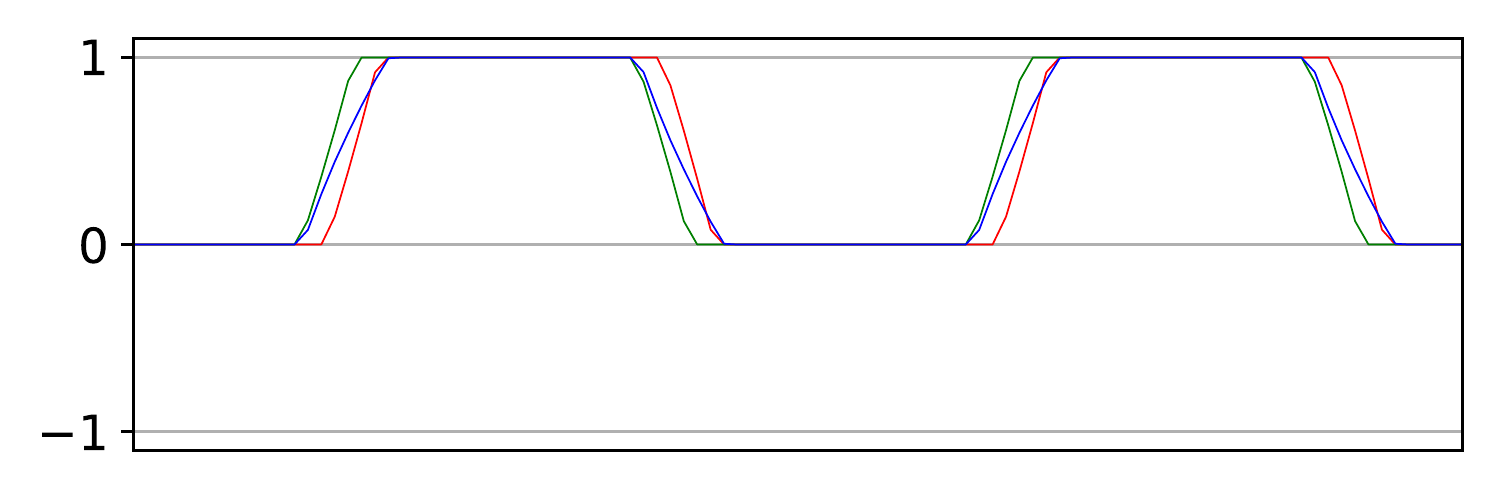}
    \end{subfigure} &
    \begin{subfigure}[b]{0.24\textwidth}
        \centering
        \includegraphics[width=\textwidth]{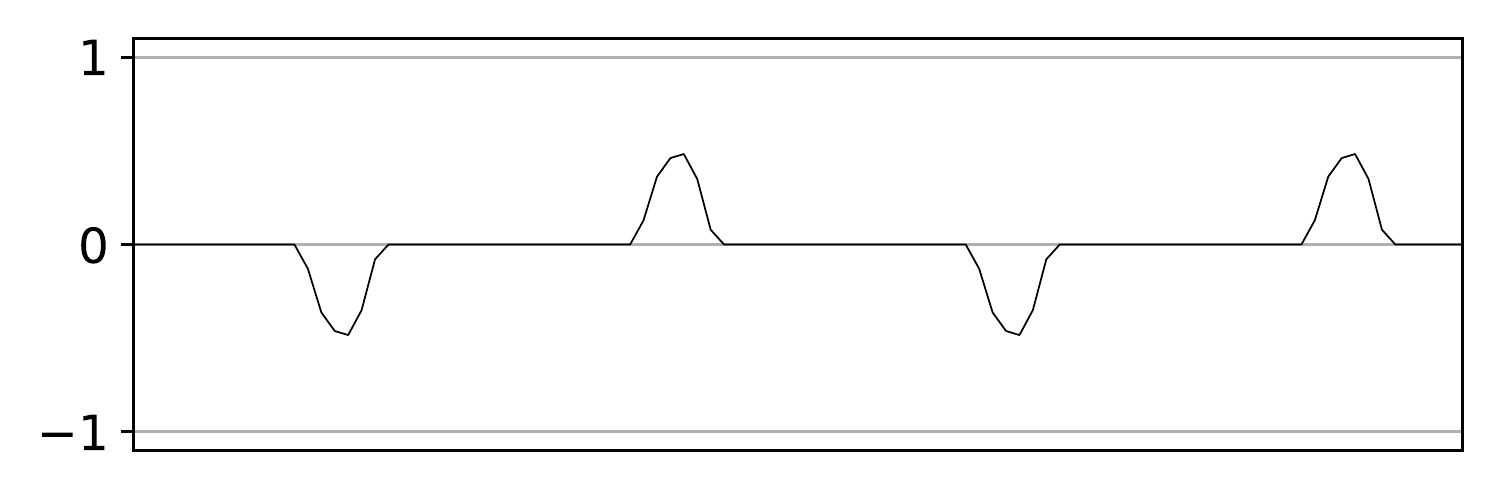}
    \end{subfigure} & 
    \begin{subfigure}[b]{0.24\textwidth}
        \centering
        \includegraphics[width=\textwidth]{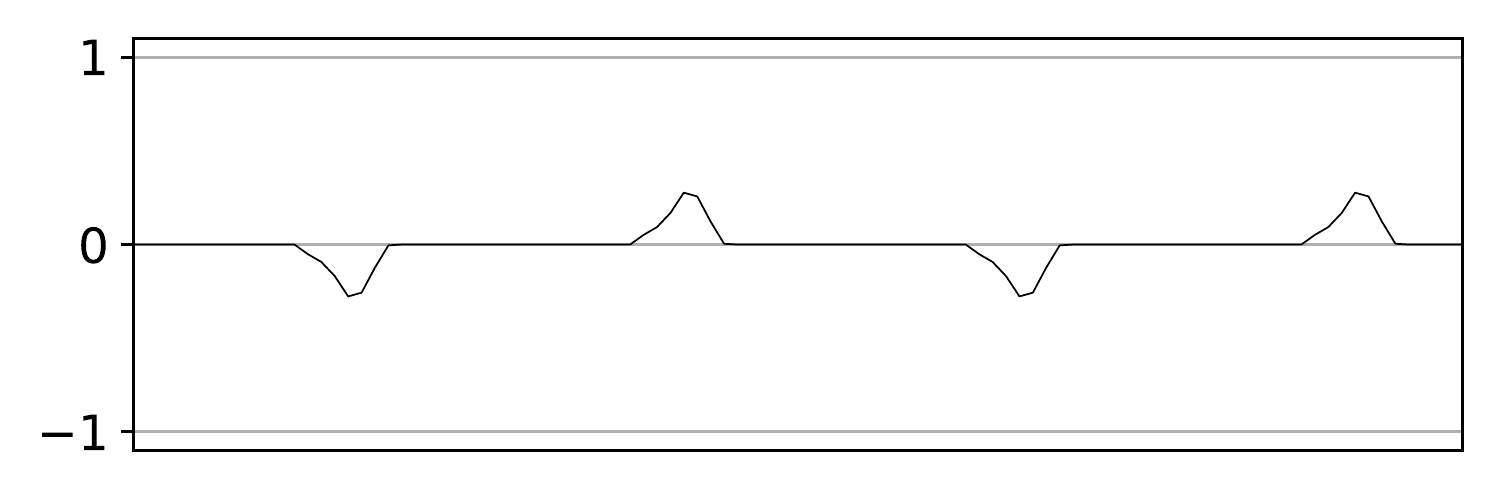}
     \end{subfigure}
     \end{tabular}
    \caption{From left to right: A synthetic aberrated image $v$, a sharpened version $z$ with noticeable 
    lateral chromatic aberration, the differences $z_R - z_G$ and $z_B - z_G$
    showing the support of the colored artifacts. 
    The profile of $z$ shows that the miscorrelation of the three color channels causes
    the remaining artifacts, detected in the profiles of the images $z_R-z_G$ and $z_B-z_G$.}
    \label{fig:colordiff}
\end{figure}

As we said in the main paper, we follow Chang \etal~\cite{chang13correction} and leverage the property that
the green/red and green/blue image residuals are good features
to detect chromatic aberrations in a photograph.
Figure \ref{fig:colordiff} shows an example for a checker grid
image. Bumps on the profiles of the residuals indicate the presence
of colored edges, most likely aberrations. When training a CNN,
we minimize these quantities so the bumps are as small as possible.

\subsection{Architecture}
We detail the architecture of $\phi$ for predicting the colored
residual in Table~\ref{tab:architecture}. We call $C$ a convolutional
layer, $R$ a ReLU activation, $B$ a batch normalization module and ``Add $d$''
a block that adds to the current feature map the output of layer $d$.
All the convolutions have $3\times3$ filters and the dimensions are given
with the format ``output/input'' channels.

Note that the input channel width is set
to 2 since it combines the green and
either the red or the blue channel, and
returns a residual for the red or
the blue channel.

\begin{table}[t]
    \centering
    \adjustbox{width=0.99\textwidth}{
    \small{
    \begin{tabular}{lcccccccccccc}
        \toprule
        Tag & 1 & 2 & 3 & 4 & 5 & 6 & 7 & 8 & 9 & 10 & 11 \\
        \midrule
        Layer & CBR & CBR & CBR & CBR & CBR & Add 3 & CBR & Add 2 & CBR & Add 1 & C\\
        Dim & $16\times2$ & $32\times16$ & $64\times32$ & $64\times 64$ & $64\times64$ & - & $32\times64$ & - & $16\times32$ & - & $1\times16$\\
        \bottomrule
    \end{tabular}
    }
    }
    \caption{Detail of the architecture of $\phi$ in the main paper for edge correction.}
    \label{tab:architecture}
\end{table}

\subsection{Training data generation}
We detail in this section the training data for learning the
optimal parameter $\nu$ of the CNN. The generation
may be divided into four main stages, resulting
in a deblurred but with colored-edges image and its sharp
counterpart:
\begin{enumerate}
    \item Unprocessing a JPEG image with the pipeline of
    Brooks \etal\cite{brooks19unprocessing}. We invert 
    tone-mapping with their proposed inverse S-curve function
    and gamma compression with the exponent $2.2$. This yields
    a synthetic RGB image with linear values with respect to the 
    electron counts;
    \item Blurring and adding noise to the raw image. The simulated blurs are Gaussian filters with standard deviation values $\rho_c$ and $\sigma_c$ $(c=R,G,B)$ in [0.2, 4] and
    sub-pixel horizontal and vertical translations in [-4,4].
    The blurry and noisy patch is finally mosaicked with the RGGB Bayer pattern;
    \item Denoising and demosaicking the synthetic raw image to mimic the two first stages of an image editing software. Because of speed for generating 
    the training data, we use the bilateral filter \cite{tomasi1998bilateral} for denoising and the Hamilton-Adams algorithm \cite{hamilton97interpolation} for demosaicking;
    \item Deblurring the raw patch with the blind deblurring
    technique detailed in Section 4.1: From the denoised
    and demosaicked patch, we first predict the
    orientation $\theta$ and the color-dependent standard
    deviation values $\sigma_c$ and $\rho_c$, and second
    we remove the blur with the approximate inverse filter
    defined by the polynomial $p$.
\end{enumerate}
In this work we used the bilateral filter of 
for denoising and the Hamilton-Adams interpolator for fast demosaicking.
We used these algorithms since they are fast but training may be indeed
enhanced with CNN-based algorithms, \eg the blind denoiser of
Wang \etal\cite{wang20practical} and the demosaicking module of 
Gharbi \etal\cite{gharbi16deep}.

\begin{figure}[t]
     \centering
     \begin{tabular}{ccc}
     \begin{subfigure}[b]{0.32\textwidth}
         \centering
         \includegraphics[width=\textwidth]{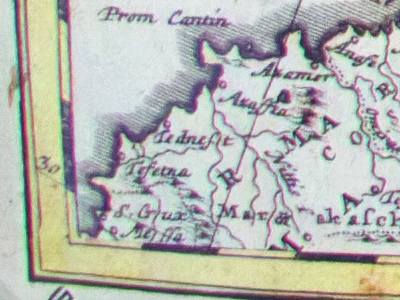}
         \caption{Aberrated.}
     \end{subfigure}  &  
        \begin{subfigure}[b]{0.32\textwidth}
         \centering
         \includegraphics[width=\textwidth]{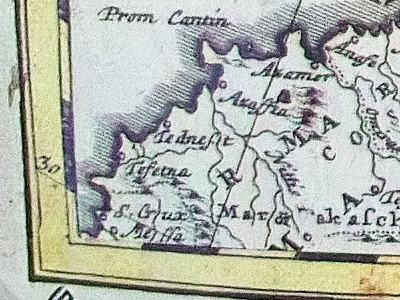}
         \caption{Loss in Eq.~(8).}
     \end{subfigure}  &  
    \begin{subfigure}[b]{0.32\textwidth}
         \centering
         \includegraphics[width=\textwidth]{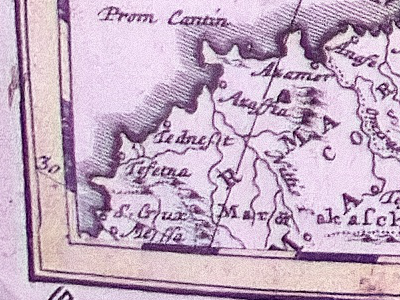}
         \caption{Gradient variant.}
     \end{subfigure} \\
     \end{tabular}
     \caption{Comparison of the color biases introduces by training loss, whether we
     evaluate the difference of the pixel values,
     \ie $\| (\widehat{u}_c - z_G) - (u_c - u_G)\|_1$,
     or the gradient values, \ie $\| \nabla(\widehat{u}_c - z_G) - \nabla(u_c - u_G)\|_1$,
     as advised by the
     prior of Heide \etal\cite{heide13simplelens}.
     The loss on pixel value residuals retains the
     same exposure and color palette as in the original aberrated image, whereas
     the one on the gradients introduces
     a pinkish bias. Note that both versions
     actually compensate the colored edges.}
     \label{fig:losscolorederror}
\end{figure}

\subsection{Choice of the loss}
We have shown in the main paper that the proposed loss in Eq.~(8),
built over red/green and blue/green residuals is pivotal to 
achieve colored edge correction.
However the prior of Heide \etal\cite{heide13simplelens} in Eq.~(13)
compares the gradients of the color channel. Training the CNN $\phi$ with 
a loss minimizing the gradients of the residuals instead, and reminiscent of 
Eq.~(13), is sub-optimal since there is no reference to the pixel, and leads to
a wrong average color in the image. Thus, the residual-based training loss
prevents these issues, leverages the property of the lateral
chromatic aberrations detailed in \cite{chang13correction}, and
leads to solutions minimizing the prior of Heide \etal at the same time.
Figure \ref{fig:losscolorederror} shows an example of this phenomenon.
We have noted that the combination of the loss on the pixel values in Eq.~(8)
and on the gradients of the color residuals was leading to marginal gains
compared to that only minimizing the pixel values of the colored residuals,
validating Eq.~(8).

\begin{figure}[t]
     \centering
     \resizebox{\textwidth}{!}{
     \begin{tabular}{ccccc}
        \begin{subfigure}[b]{0.2\textwidth}
         \centering
         \includegraphics[width=\textwidth]{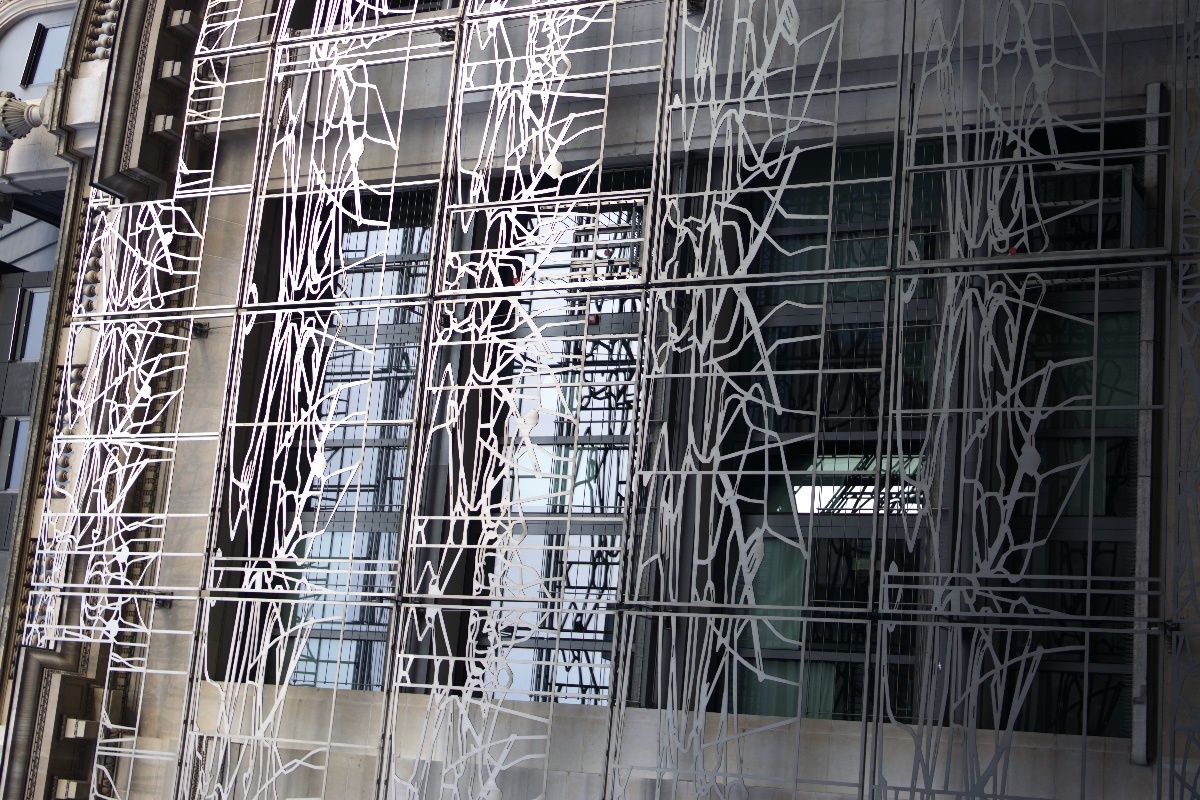}
     \end{subfigure}  &  
    \begin{subfigure}[b]{0.2\textwidth}
         \centering
         \includegraphics[width=\textwidth]{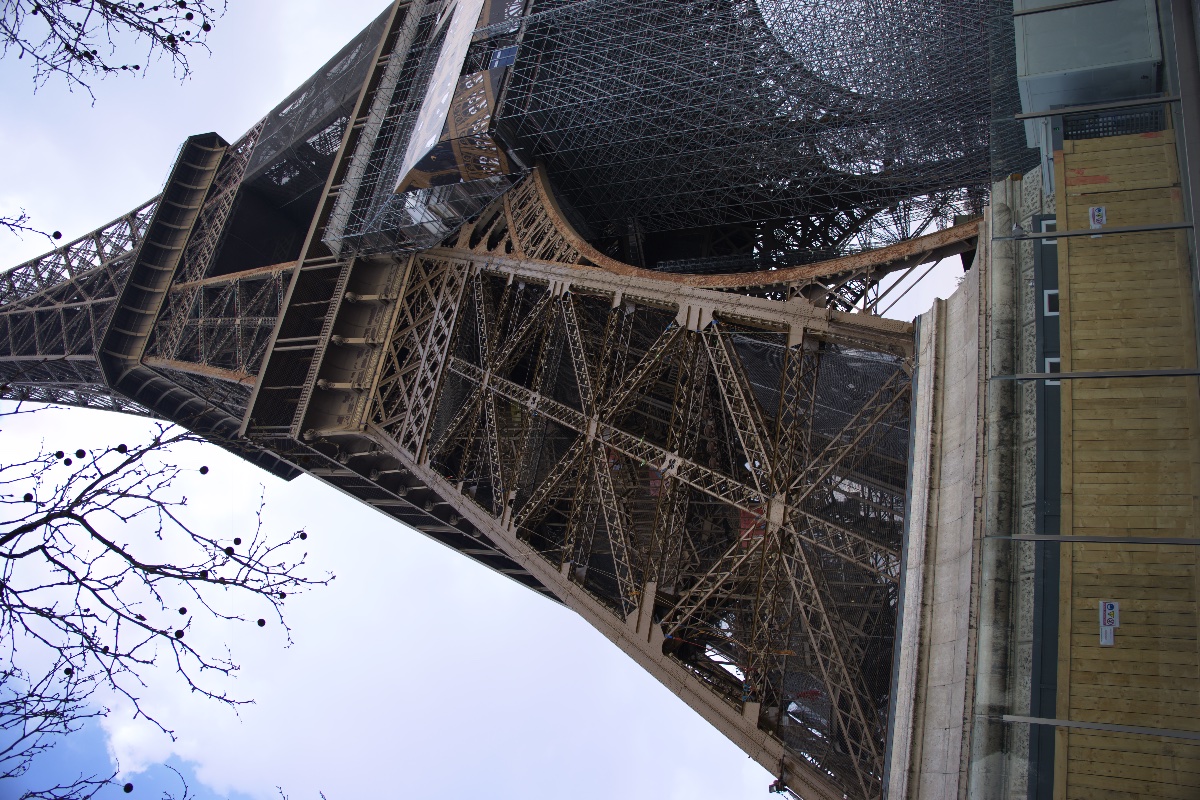}
     \end{subfigure} &
    \begin{subfigure}[b]{0.2\textwidth}
         \centering
         \includegraphics[width=\textwidth]{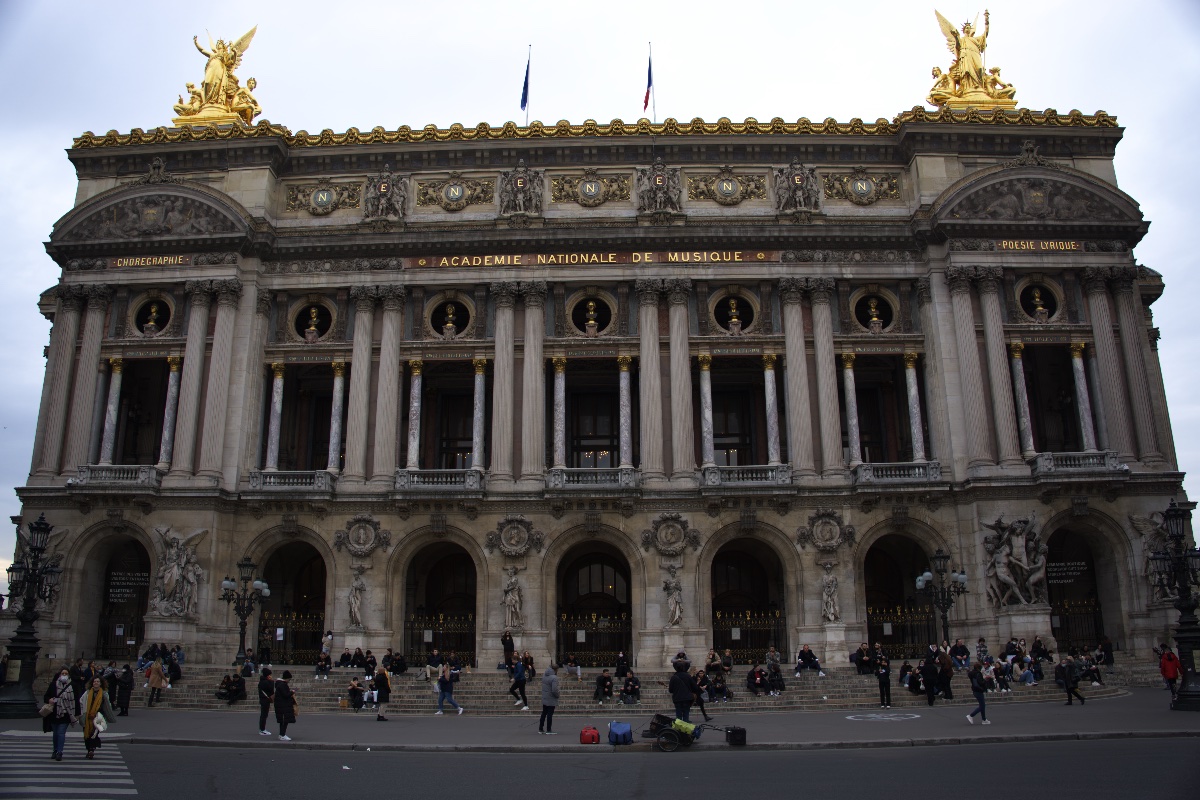}
     \end{subfigure} &
    \begin{subfigure}[b]{0.2\textwidth}
         \centering
         \includegraphics[width=\textwidth]{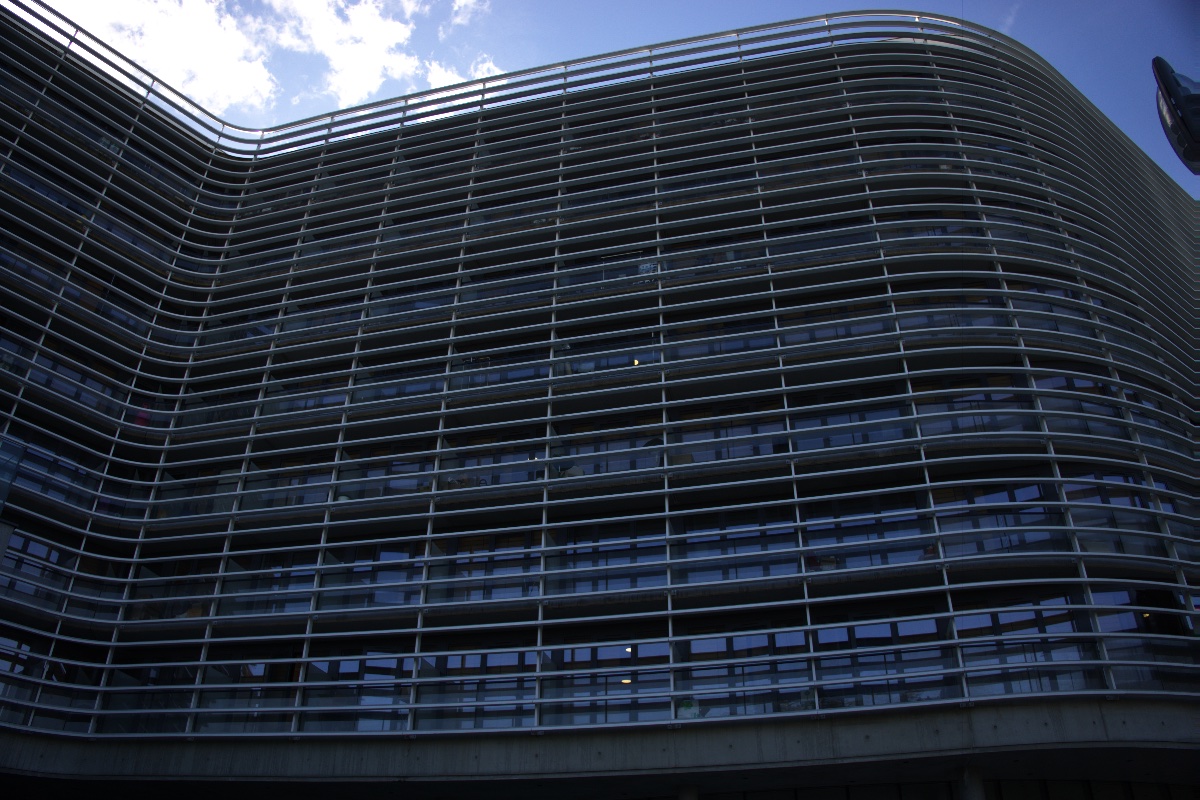}
     \end{subfigure}  &  
    \begin{subfigure}[b]{0.2\textwidth}
         \centering
         \includegraphics[width=\textwidth]{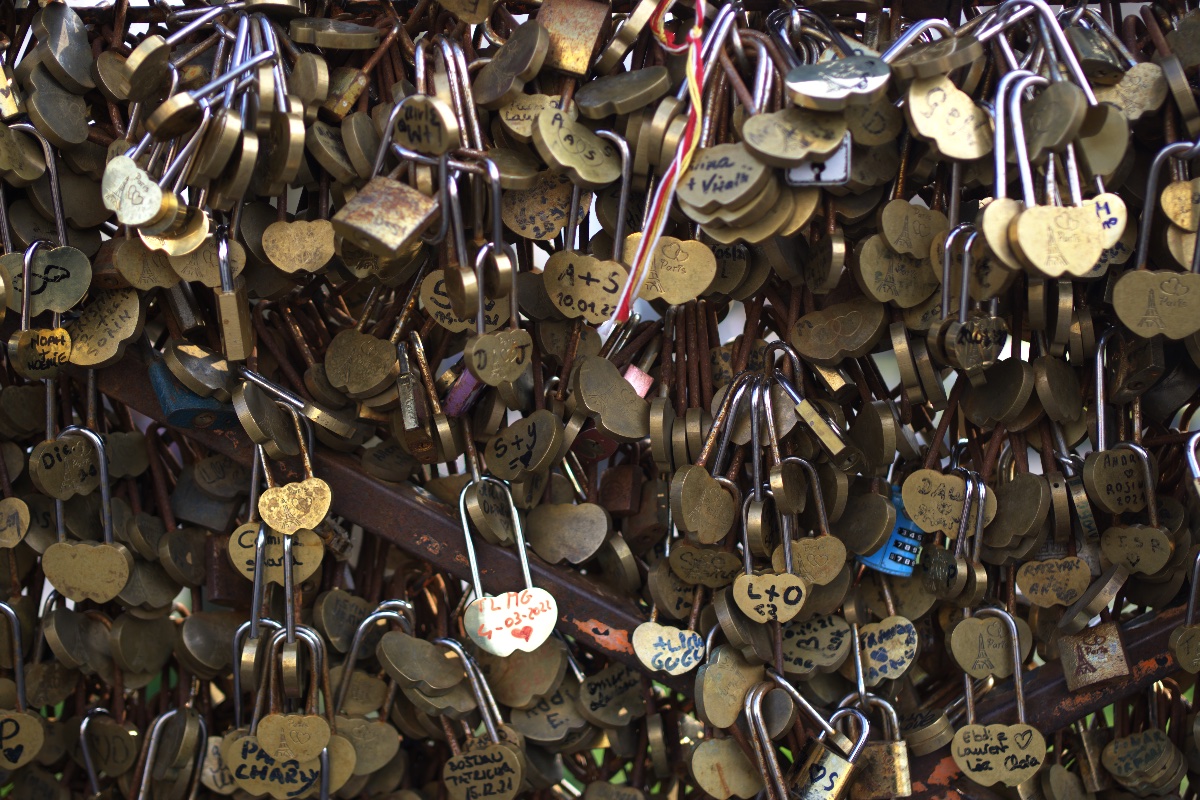}
     \end{subfigure} \\
    \begin{subfigure}[b]{0.2\textwidth}
         \centering
         \includegraphics[width=\textwidth]{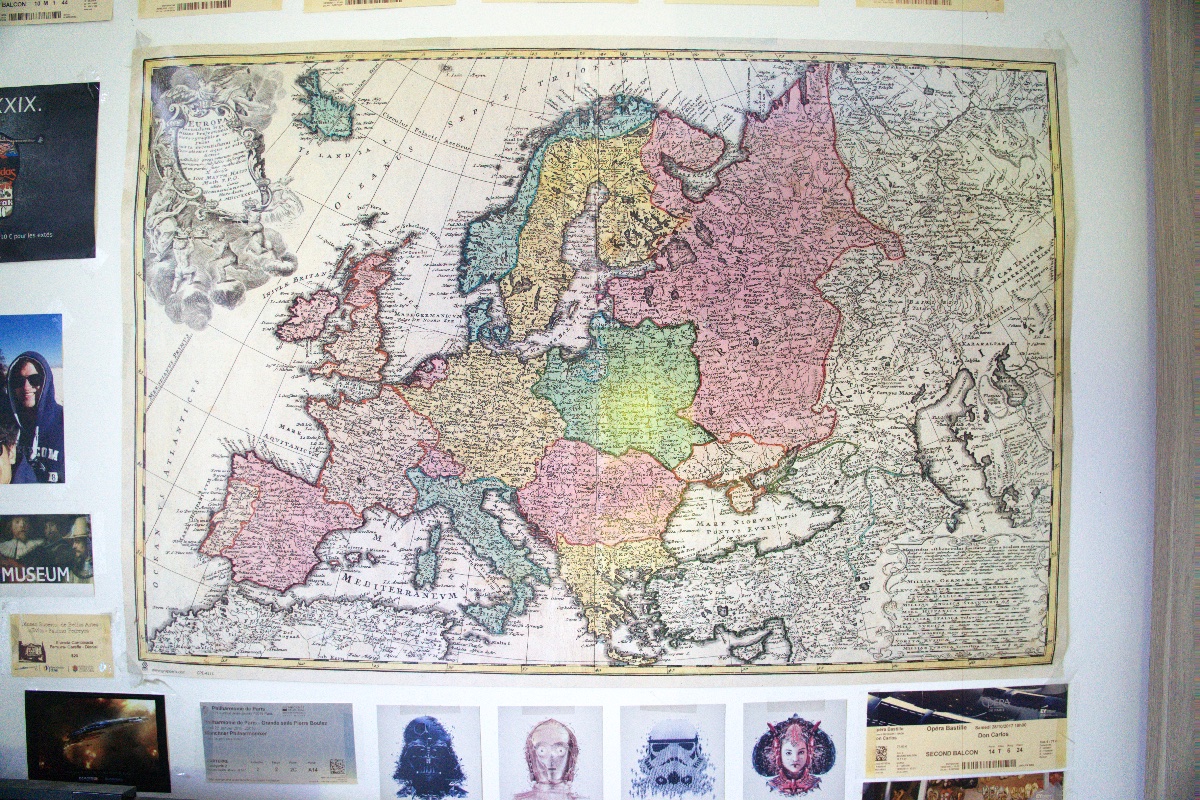}
     \end{subfigure} & 
    \begin{subfigure}[b]{0.2\textwidth}
         \centering
         \includegraphics[width=\textwidth]{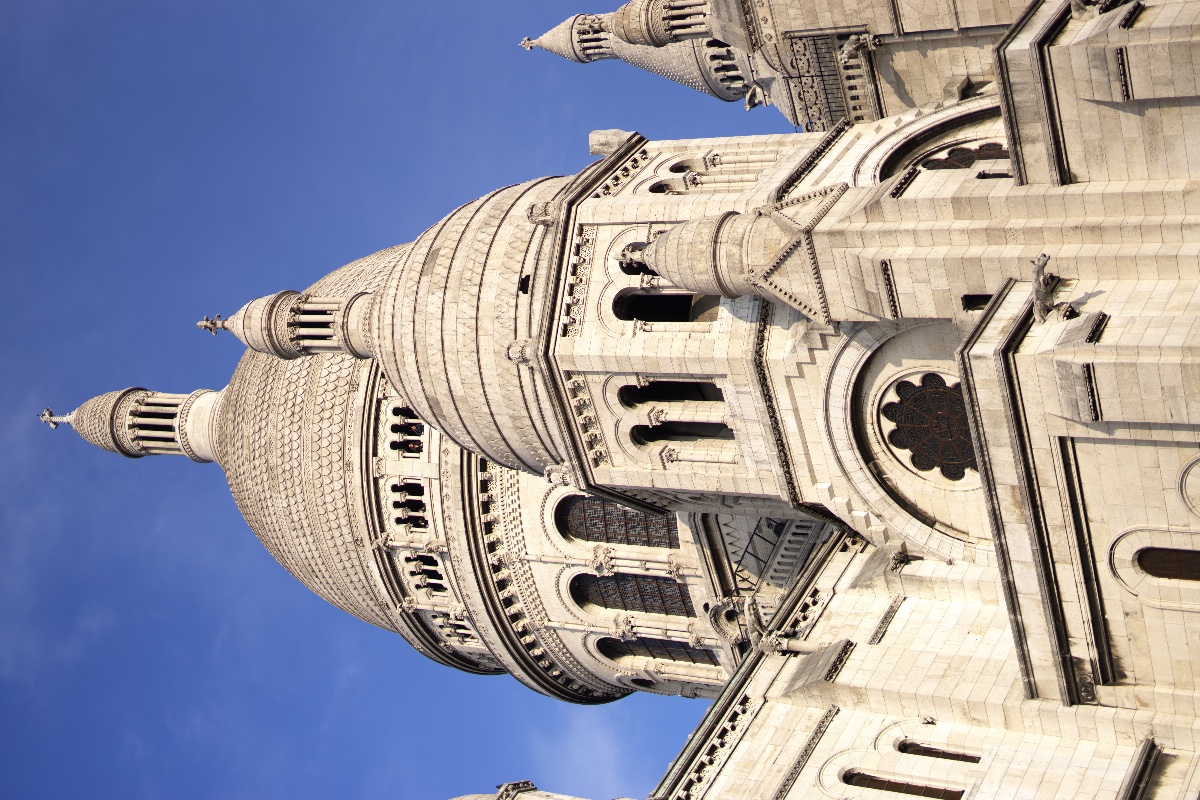}
     \end{subfigure} & 
    \begin{subfigure}[b]{0.2\textwidth}
         \centering
         \includegraphics[width=\textwidth]{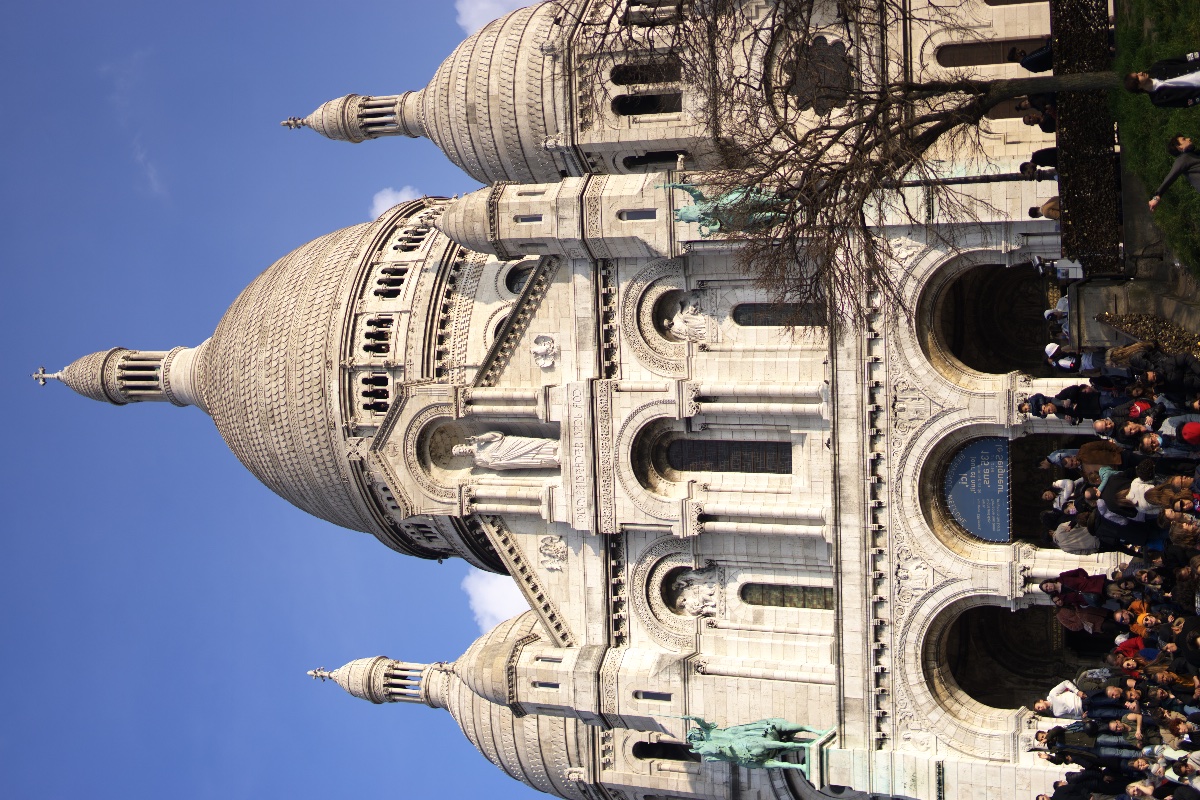}
     \end{subfigure} & 
    \begin{subfigure}[b]{0.2\textwidth}
         \centering
         \includegraphics[width=\textwidth]{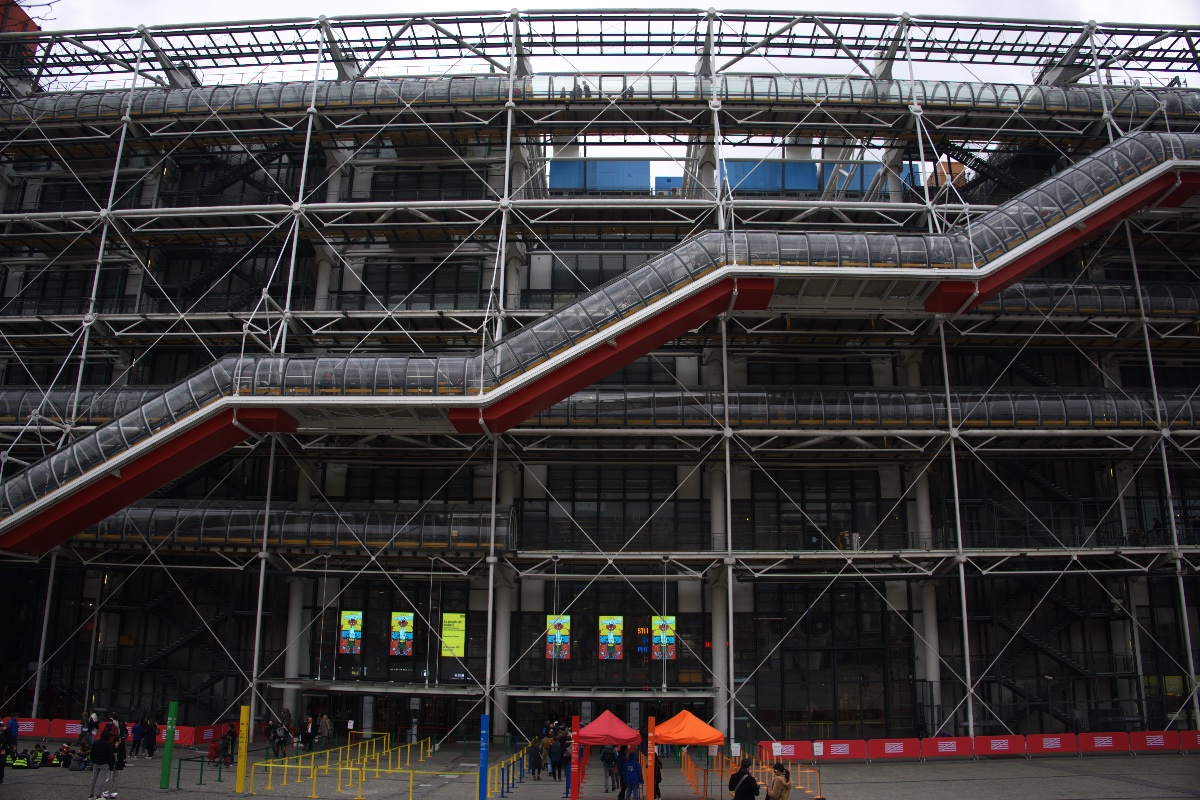}
     \end{subfigure} & 
    \begin{subfigure}[b]{0.2\textwidth}
         \centering
         \includegraphics[width=\textwidth]{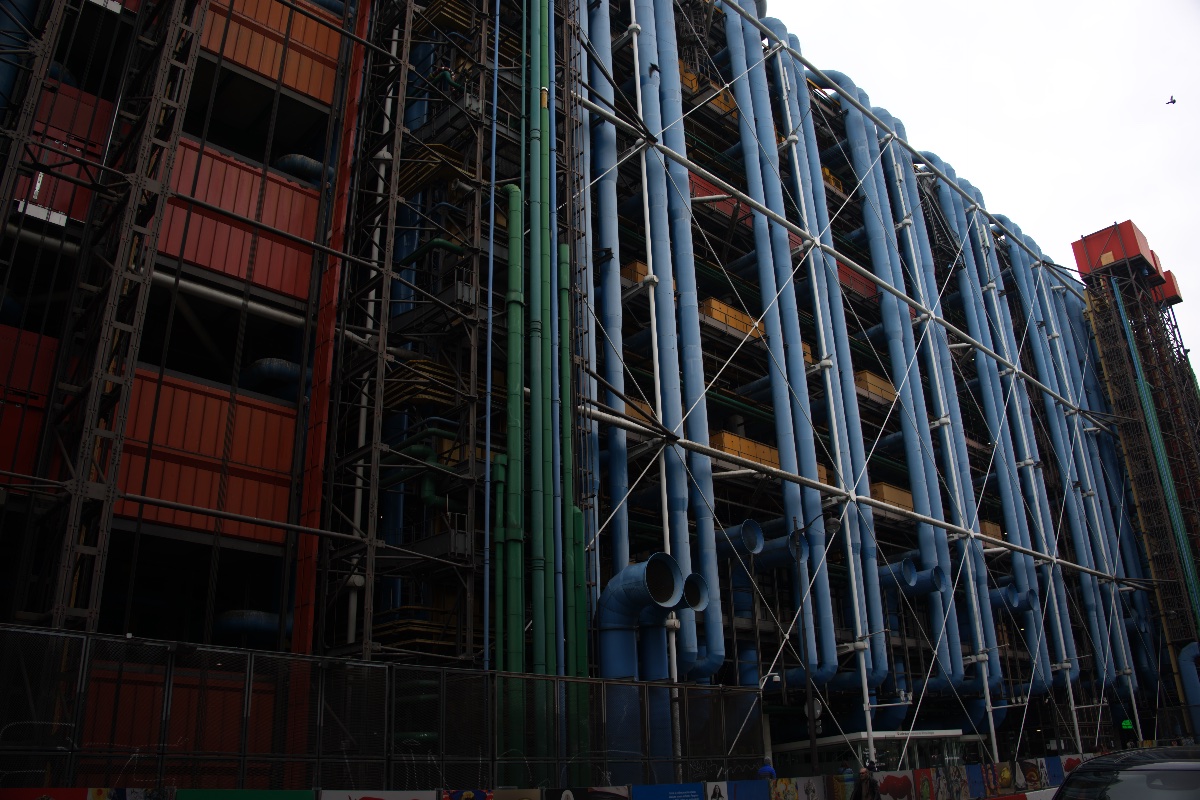}
     \end{subfigure} \\
     \end{tabular}     
     }
    \caption{The ten $6000\times4000$ images (24 megapixels) we use
    for the quantitative analysis of the edge correction algorithm. Each image features salient edges, prone to lateral chromatic
    aberrations. The reader is invited to zoom in on a computer screen.}
    \label{fig:benchmark}
\end{figure}

\section{Whole pipeline details}
\label{sec:pipeline}

\subsection{Test images}
We show in Figure~\ref{fig:benchmark} the ten images we use for
evaluating the edge correction algorithms in the experiment section
of the main paper. The images where taken with the Sony $\alpha6000$
camera, the Sony FE 35mm $f/1.8$ lens at maximal aperture
and the Sigma 18-50mm $f/2.8$ DC DN lens at maximal aperture and shortest
focal length.

\begin{figure}[t]
     \centering
     \begin{tabular}{ccccc}
     \begin{subfigure}[b]{0.19\textwidth}
         \centering
         \includegraphics[width=\textwidth]{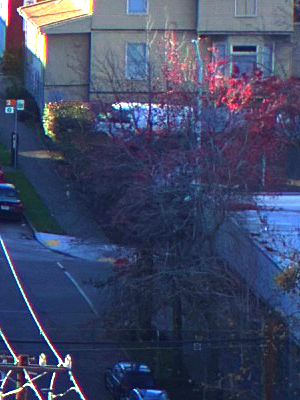}
         \caption{Original}
     \end{subfigure}  &  
        \begin{subfigure}[b]{0.19\textwidth}
         \centering
         \includegraphics[width=\textwidth]{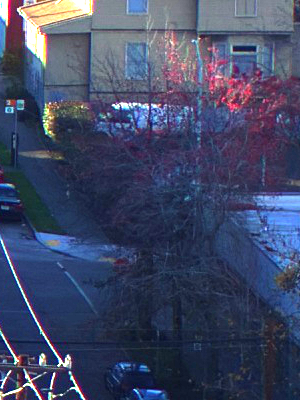}
         \caption{Size 100}
     \end{subfigure}  &  
    \begin{subfigure}[b]{0.19\textwidth}
         \centering
         \includegraphics[width=\textwidth]{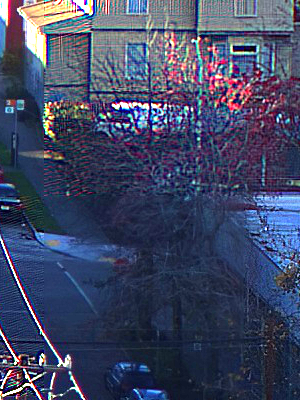}
         \caption{Size 200}
     \end{subfigure} &
    \begin{subfigure}[b]{0.19\textwidth}
         \centering
         \includegraphics[width=\textwidth]{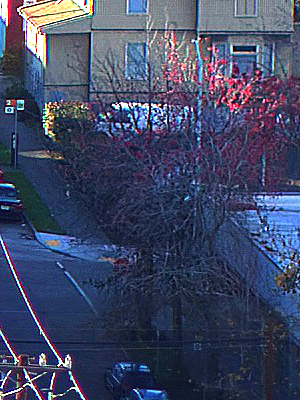}
         \caption{Size 400}
     \end{subfigure} &
    \begin{subfigure}[b]{0.19\textwidth}
         \centering
         \includegraphics[width=\textwidth]{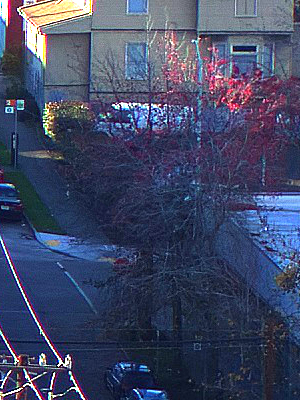}
         \caption{Size 800}
     \end{subfigure} \\
     \end{tabular}
     \caption{Qualitative comparison of the impact of the patch size on 
     the peformance of the blind deblurring module on a real aberrated image. 
     From left to right: 
     the aberrated image and restored versions where the patch size 
     is respectively 100, 200, 400 and 800.
     For patch sizes of 400 or 800, the image is actually deblurred. 
     Under, the image is either blurry or contains artifacts. Note the
     presence of colored edges, \eg next ot the electrical cable, since
     we show images solely deblurred, prior to any evaluation with the
     CNN $\phi$. The 
     reader is invited to zoom in on a computer screen.}
     \label{fig:patchsize}
\end{figure}

\subsection{Patch size}
Setting the size of the patches is critical for the success of the blur estimation
technique of Delbracio \etal\cite{delbracio21polyblur} in the context 
of spatially-varying blurs.
Indeed, this method is based on the presence of salient edges and may fail 
if there are too few edges on the patch support, \eg for too small patches.
We show in Fig.~\ref{fig:patchsize} the comparison of an image crop
for images deblurred with different patch sizes, ranging from 100 to 800 pixels.
We restore non-overlapping patches to visualize what the deblurring exactly
restores on each patch.
In this figure one can see that for the patch size set to 100 the image looks almost like the original one. For the patch size set to 200, noticeable deconvolution
artifacts can be seen next to the leaves. For the patch size set to 400 and 800, the
restored results are plausible.

The result for the patch size of 200 may be explained by the fact that, to work well the blur estimation method needs edges with important contrast.
However, in textured regions with only moderate gradients, the affine rule
may predict a larger standard deviation value than the real one, resulting
in a too large deconvolution filter and thus artifacts in the final image.
A Weakness of this affine rule is thus such regions, and a simple way to prevent
these artifacts is selecting larger patches to favor the presence of more
contrasted edges. In this presentation, we
set the patch size to 400, which is valid 
for most images we have tested our approach
on.

\begin{figure}[t]
     \centering
     \begin{tabular}{cccc}
    \begin{subfigure}[b]{0.24\textwidth}
         \centering
         \includegraphics[width=\textwidth]{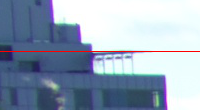}
     \end{subfigure}  &  
    \begin{subfigure}[b]{0.24\textwidth}
         \centering
         \includegraphics[width=\textwidth]{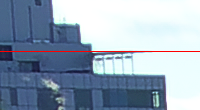}
     \end{subfigure} &
     \begin{subfigure}[b]{0.24\textwidth}
         \centering
         \includegraphics[width=\textwidth]{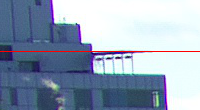}
     \end{subfigure} &
    \begin{subfigure}[b]{0.24\textwidth}
         \centering
         \includegraphics[width=\textwidth]{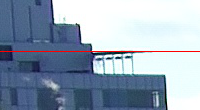}
     \end{subfigure} \\
    \begin{subfigure}[b]{0.24\textwidth}
         \centering
         \includegraphics[width=\textwidth]{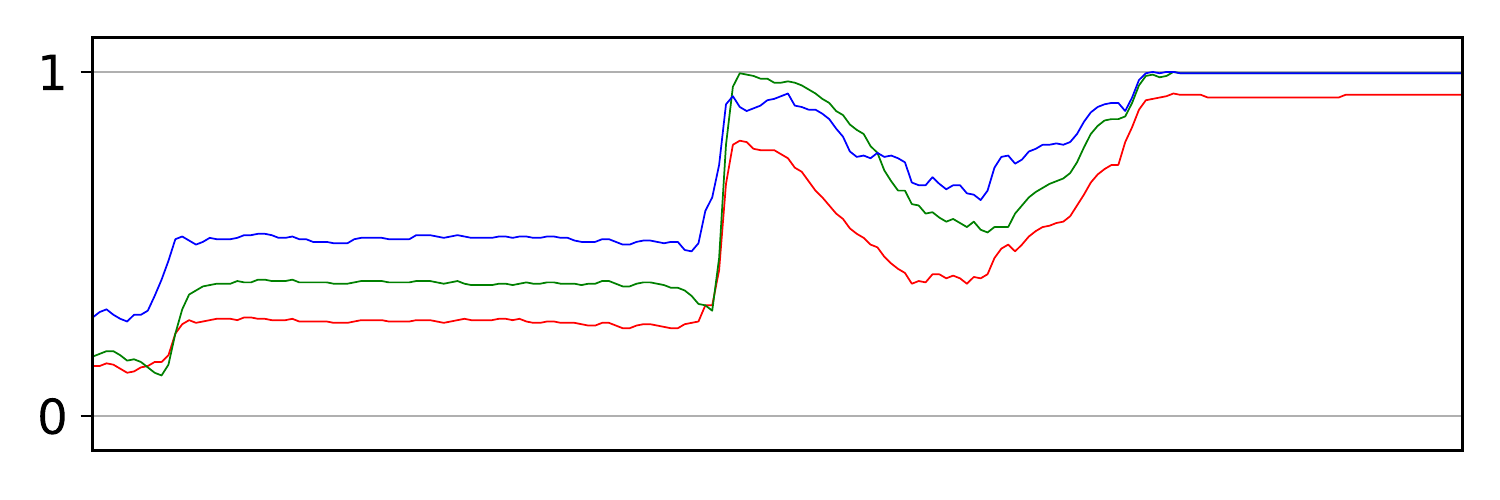}
     \end{subfigure}  &  
    \begin{subfigure}[b]{0.24\textwidth}
         \centering
         \includegraphics[width=\textwidth]{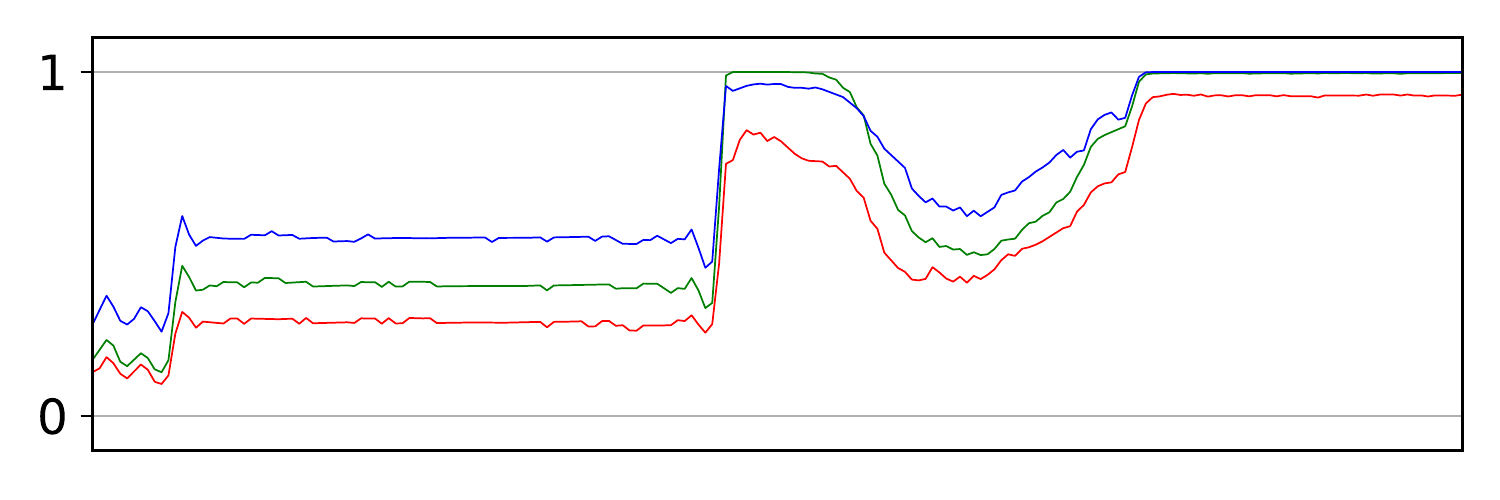}
     \end{subfigure} &
    \begin{subfigure}[b]{0.24\textwidth}
         \centering
         \includegraphics[width=\textwidth]{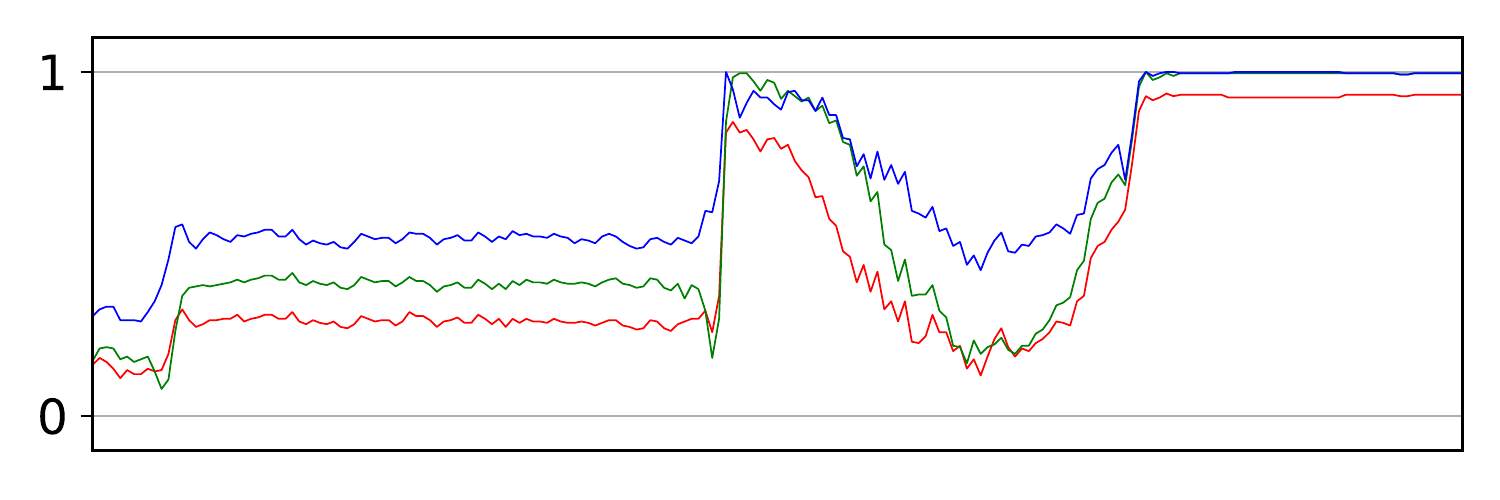}
     \end{subfigure} &
    \begin{subfigure}[b]{0.24\textwidth}
         \centering
         \includegraphics[width=\textwidth]{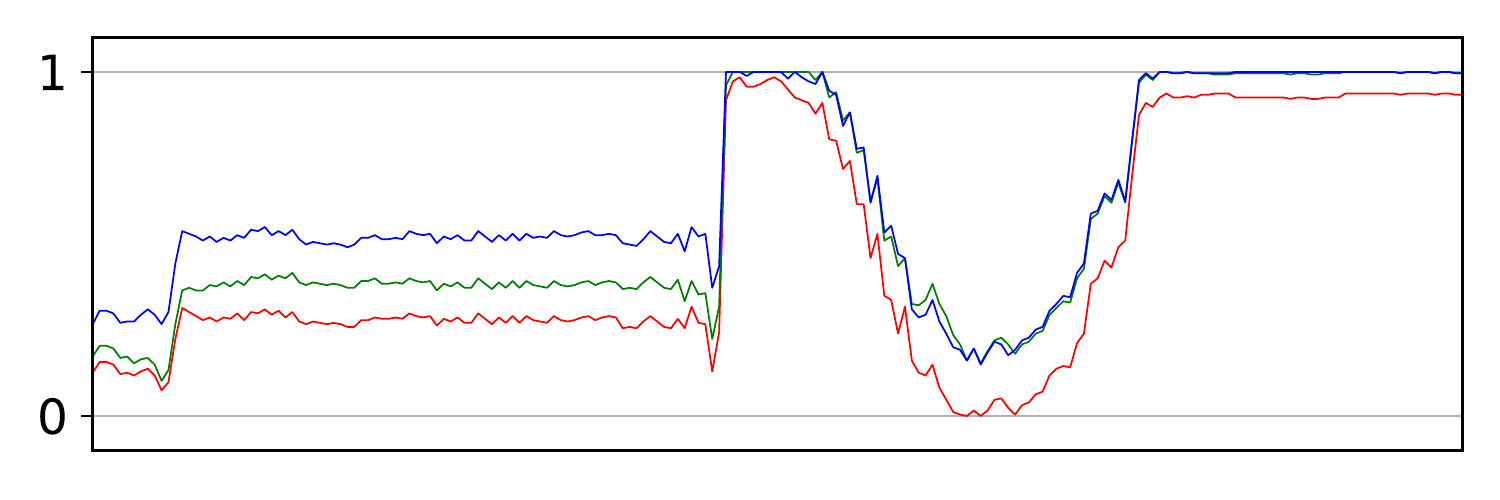}
     \end{subfigure}
     \end{tabular}
    \caption{Comparison of purple fringe removal from a real raw image. 
    From left to right: The blurry image, the 
    version restored with
    DxO PhotoLab 5 (non-blind), and the images only 
    deblurred and deblurred+corrected by our model (blind).}
    \label{fig:purplefringes}
\end{figure}

\subsection{Saturation}
The combination of optical chromatic aberrations and 
saturation is called by photographers 
``purple fringes'', an artifact challenging to remove.
Yet, our technique successfully compensates these fringes as
shown in Fig.~\ref{fig:purplefringes}.
However, despite good performance on real images our approach cannot remove 
all the purple fringes, and leaves a thin dark line next to saturated areas.
As previously noted, the performance of our method
is closely related to that of the blur estimation stage, which
makes the assumption that there is at least one strong edge in the patch,
and thus may fail in textured regions.

\subsection{Selection of the Sub-modules}
To validate the choices of Polyblur~\cite{delbracio21polyblur} and
the novel 2-channel CNN, we compare these algorithms
with typical candidates for these two stages.

Since it is fast and capture relatively well real-world PSFs,
we predict the blur kernels with the estimator of 
Delbracio~\etal\cite{delbracio21polyblur}. The first experiment
in the main paper have shown that these kernels approximate well
lens blurs. We thus compare
the approximate inverse filter from \cite{delbracio21polyblur},
suited for the predicted ``mild'' blurs beforehand, and a 
state-of-the-art CNN for non-blind image deblurring whose 
backbone is featured in \cite{li21universal}.

Second, we compensate the optical aberrations with typical
registration techniques such as global radial model solved
with a simplex algorithm~\cite{kang07automatic}, local 
translations predicted with the Lucas-Kanade 
algorithm~\cite{baker04lucas} and our CNN. Unfortunately there
is no available code for the technique of Chang~\etal\cite{chang13correction}.

\section{Additional images}
\label{sec:images}

\subsection{Raw images}
\begin{figure}[t]
    \centering
    \begin{tabular}{ccc}
     \begin{subfigure}[b]{0.32\textwidth}
         \centering
         \includegraphics[width=\textwidth]{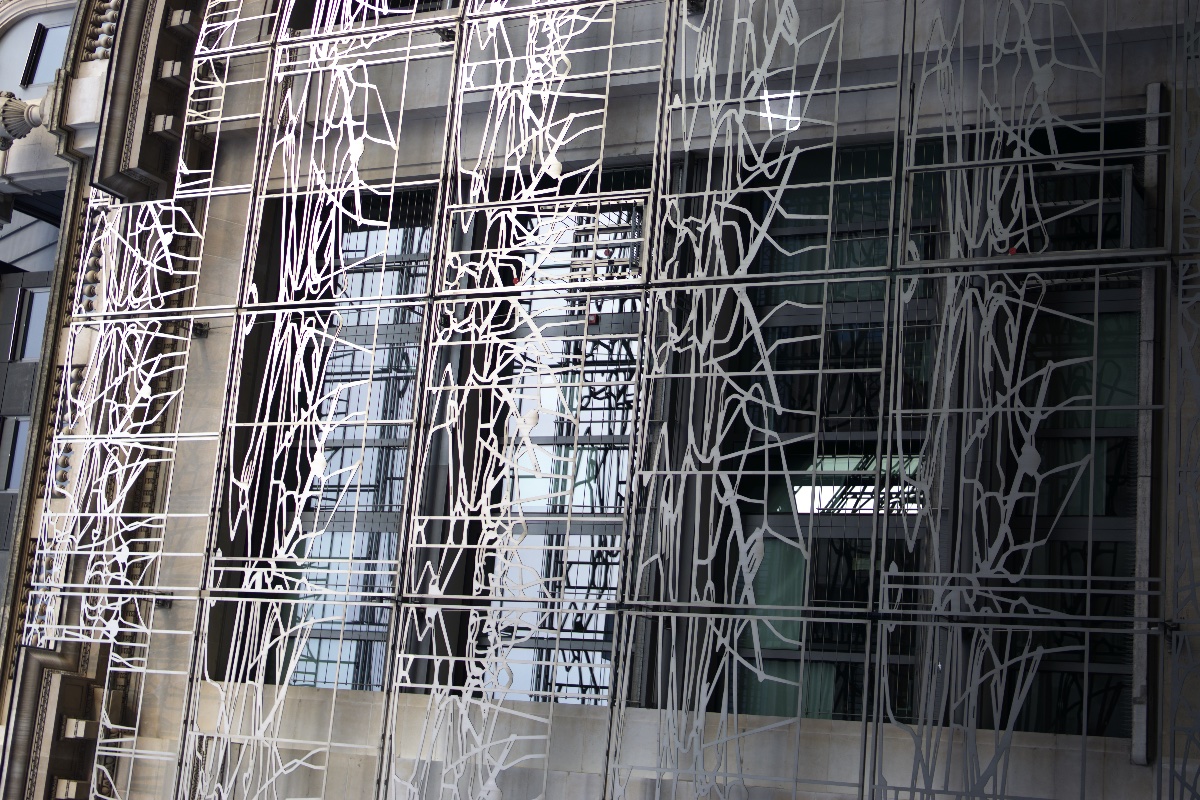}
         \caption{Culture (Sony lens).}
     \end{subfigure}  &  
        \begin{subfigure}[b]{0.32\textwidth}
         \centering
         \includegraphics[width=\textwidth]{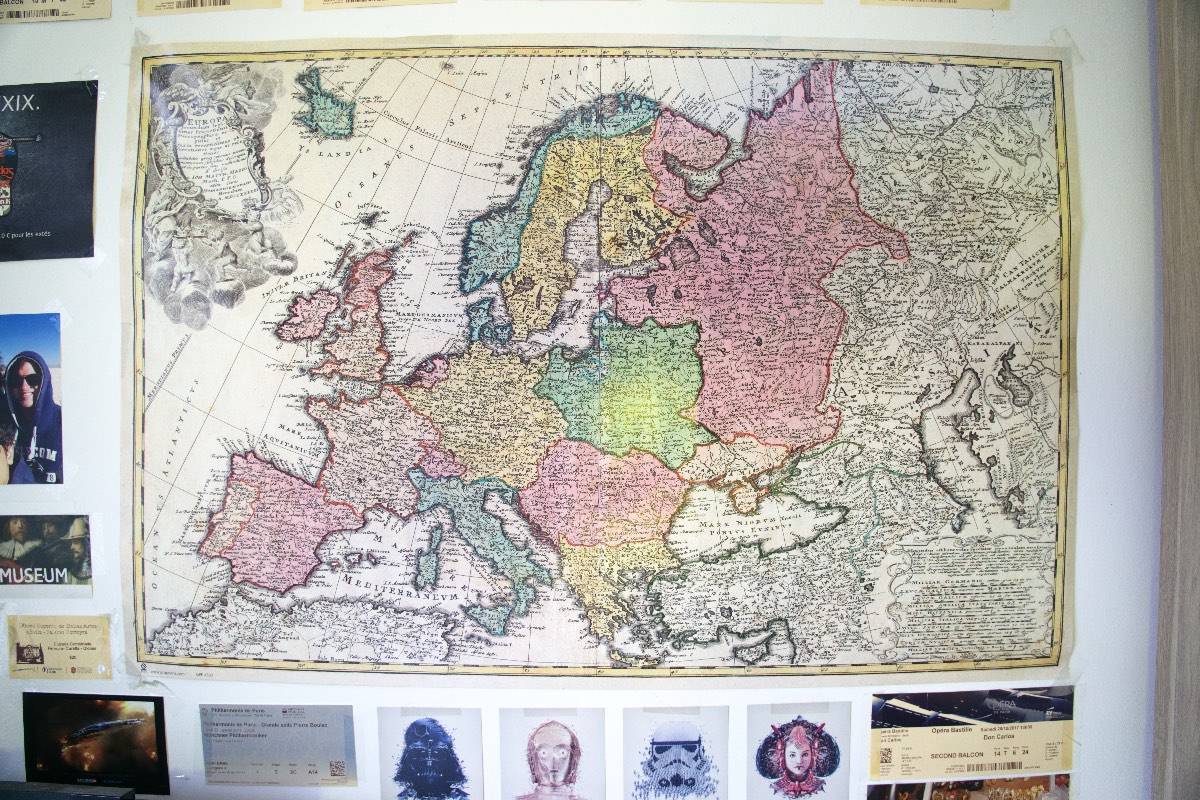}
         \caption{Map (Sigma lens).}
     \end{subfigure}  &  
    \begin{subfigure}[b]{0.32\textwidth}
         \centering
         \includegraphics[width=\textwidth]{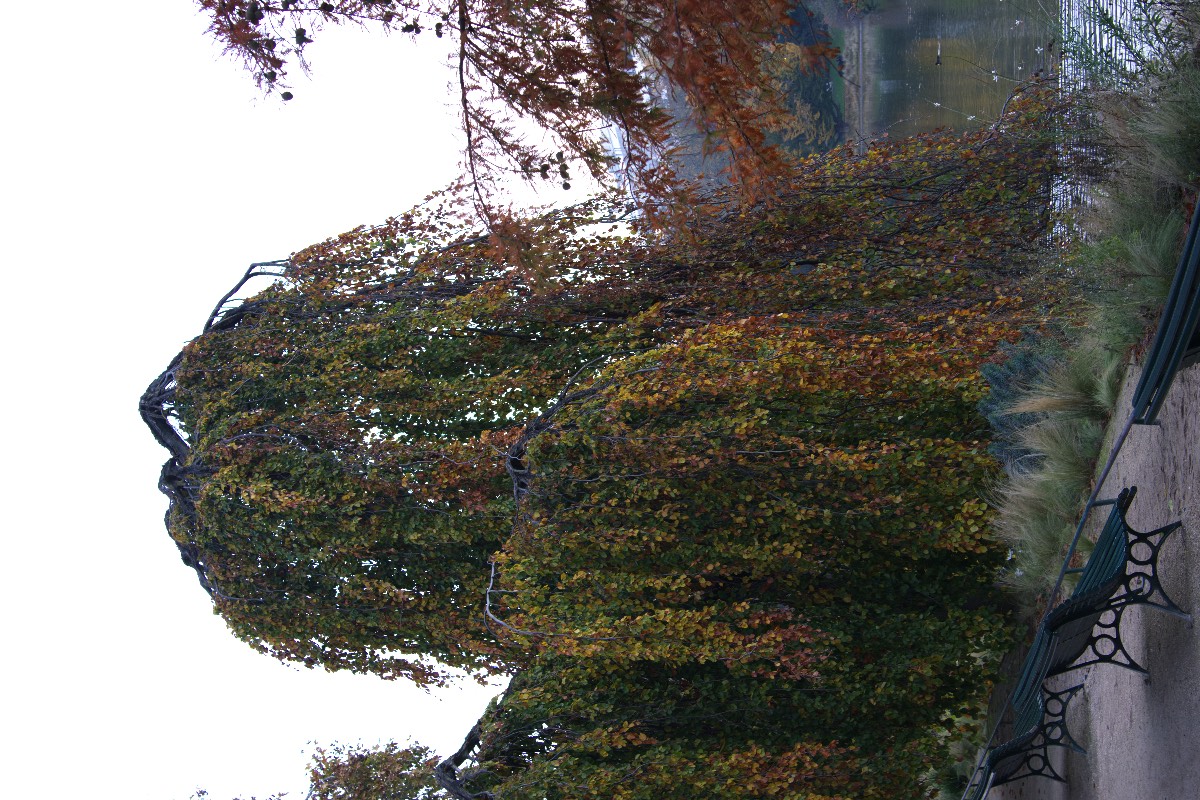}
         \caption{Tree (Sony lens).}
     \end{subfigure} \\
    \end{tabular}
    \caption{The additional images for qualitative evaluation. The images are denoised and demosaicked with DxO PhotoLab 5.}
    \label{fig:rawimages}
\end{figure}

\begin{figure}
    \centering
    \begin{tabular}{ccc}
     \begin{subfigure}[b]{0.32\textwidth}
         \centering
         \includegraphics[width=\textwidth]{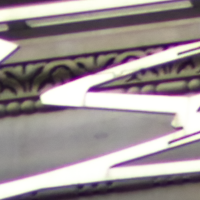}
     \end{subfigure}  &  
        \begin{subfigure}[b]{0.32\textwidth}
         \centering
         \includegraphics[width=\textwidth]{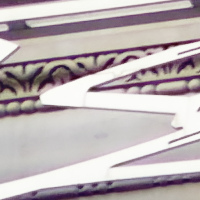}
     \end{subfigure}  &  
    \begin{subfigure}[b]{0.32\textwidth}
         \centering
         \includegraphics[width=\textwidth]{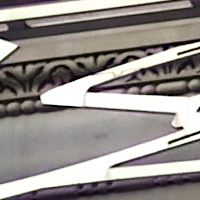}
     \end{subfigure} \\
     \begin{subfigure}[b]{0.32\textwidth}
         \centering
         \includegraphics[width=\textwidth]{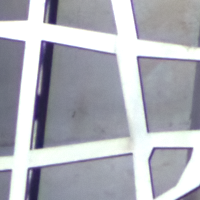}
     \end{subfigure}  &  
        \begin{subfigure}[b]{0.32\textwidth}
         \centering
         \includegraphics[width=\textwidth]{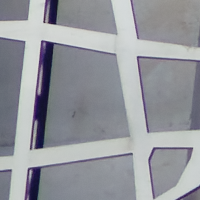}
     \end{subfigure}  &  
    \begin{subfigure}[b]{0.32\textwidth}
         \centering
         \includegraphics[width=\textwidth]{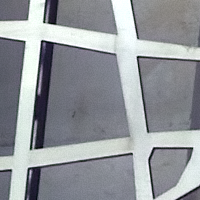}
     \end{subfigure} \\
          \begin{subfigure}[b]{0.32\textwidth}
         \centering
         \includegraphics[width=\textwidth]{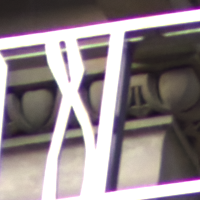}
         \caption{Blurry.}
     \end{subfigure}  &  
        \begin{subfigure}[b]{0.32\textwidth}
         \centering
         \includegraphics[width=\textwidth]{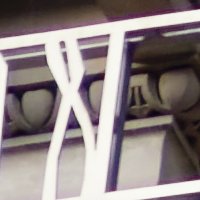}
         \caption{DxO.}
     \end{subfigure}  &  
    \begin{subfigure}[b]{0.32\textwidth}
         \centering
         \includegraphics[width=\textwidth]{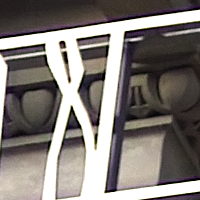}
         \caption{Ours.}
     \end{subfigure} \\
    \end{tabular}
    \caption{Crops for the ``map'' image taken with the Sony FE 35mm $f/1.8$ lens. From left to right: the original blurry image, the optical aberration correction
    of DxO PhotoLab (non-blind setting), and ours.}
    \label{fig:culture}
\end{figure}

\begin{figure}
    \centering
    \begin{tabular}{ccc}
     \begin{subfigure}[b]{0.32\textwidth}
         \centering
         \includegraphics[width=\textwidth]{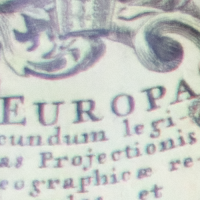}
     \end{subfigure}  &  
        \begin{subfigure}[b]{0.32\textwidth}
         \centering
         \includegraphics[width=\textwidth]{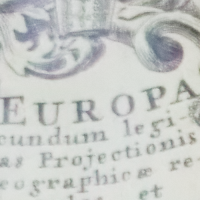}
     \end{subfigure}  &  
    \begin{subfigure}[b]{0.32\textwidth}
         \centering
         \includegraphics[width=\textwidth]{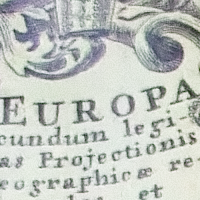}
     \end{subfigure} \\
     \begin{subfigure}[b]{0.32\textwidth}
         \centering
         \includegraphics[width=\textwidth]{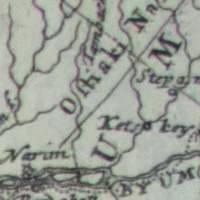}
     \end{subfigure}  &  
        \begin{subfigure}[b]{0.32\textwidth}
         \centering
         \includegraphics[width=\textwidth]{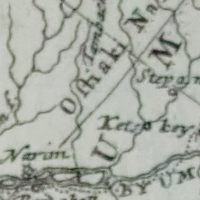}
     \end{subfigure}  &  
    \begin{subfigure}[b]{0.32\textwidth}
         \centering
         \includegraphics[width=\textwidth]{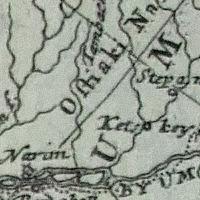}
     \end{subfigure} \\
          \begin{subfigure}[b]{0.32\textwidth}
         \centering
         \includegraphics[width=\textwidth]{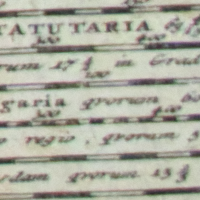}
         \caption{Blurry.}
     \end{subfigure}  &  
        \begin{subfigure}[b]{0.32\textwidth}
         \centering
         \includegraphics[width=\textwidth]{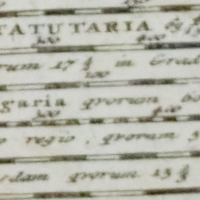}
         \caption{DxO.}
     \end{subfigure}  &  
    \begin{subfigure}[b]{0.32\textwidth}
         \centering
         \includegraphics[width=\textwidth]{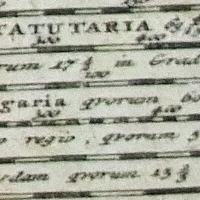}
         \caption{Ours.}
     \end{subfigure} \\
    \end{tabular}
    \caption{Crops for the ``map'' image taken with the Sigma 18-50mm $f/2.8$ DC DN lens. From left to right: the original blurry image, the optical aberration correction
    of DxO PhotoLab (blind setting), and ours.}
    \label{fig:map}
\end{figure}

\begin{figure}
    \centering
    \begin{tabular}{ccc}
     \begin{subfigure}[b]{0.32\textwidth}
         \centering
         \includegraphics[width=\textwidth]{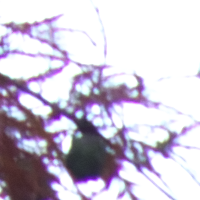}
     \end{subfigure}  &  
        \begin{subfigure}[b]{0.32\textwidth}
         \centering
         \includegraphics[width=\textwidth]{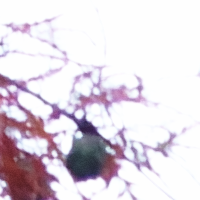}
     \end{subfigure}  &  
    \begin{subfigure}[b]{0.32\textwidth}
         \centering
         \includegraphics[width=\textwidth]{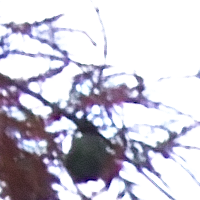}
     \end{subfigure} \\
     \begin{subfigure}[b]{0.32\textwidth}
         \centering
         \includegraphics[width=\textwidth]{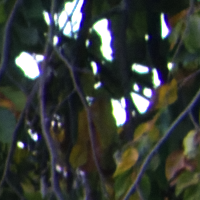}
     \end{subfigure}  &  
        \begin{subfigure}[b]{0.32\textwidth}
         \centering
         \includegraphics[width=\textwidth]{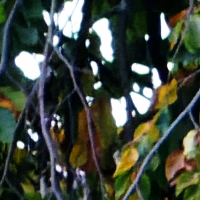}
     \end{subfigure}  &  
    \begin{subfigure}[b]{0.32\textwidth}
         \centering
         \includegraphics[width=\textwidth]{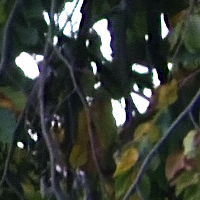}
     \end{subfigure} \\
          \begin{subfigure}[b]{0.32\textwidth}
         \centering
         \includegraphics[width=\textwidth]{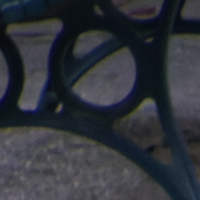}
         \caption{Blurry.}
     \end{subfigure}  &  
        \begin{subfigure}[b]{0.32\textwidth}
         \centering
         \includegraphics[width=\textwidth]{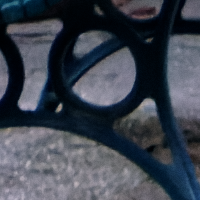}
         \caption{DxO.}
     \end{subfigure}  &  
    \begin{subfigure}[b]{0.32\textwidth}
         \centering
         \includegraphics[width=\textwidth]{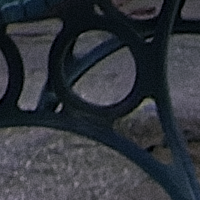}
         \caption{Ours.}
     \end{subfigure} \\
    \end{tabular}
        \caption{Crops for the ``tree'' image taken with the Sony FE 35mm $f/1.8$ lens. From left to right: the original blurry image, the optical aberration correction
    of DxO PhotoLab (non-blind setting), and ours.}
    \label{fig:tree}
\end{figure}

To qualitatively validate our approach, we have taken several photographs
with a Sony $\alpha$6000 camera, and combined with the Sony FE 35mm $f/1,8$
and the Sigma 18-50mm $f/2.8$ DC DN lenses. We compare our approach
to the commercial software DxO PhotoLab 5, whose catalog contains the profile
of the Sony lens, but not that of the Sigma one recently released in October 2021.
PhotoLab thus runs in a non-blind setting for the 35mm lens, and should achieve
the best result over our technique, whereas it runs in a blind setting for the
Sigma lens. The images dubbed ``culture'', ``map'' and
``tree'' are shown in Figure \ref{fig:rawimages},
and magnified crops are shown in Figures \ref{fig:culture}, \ref{fig:map} and \ref{fig:tree}.
The tree example in Figure~\ref{fig:tree} illustrates in particular the robustness
of our method to ``purple fringes'', \ie the combination of optical
aberrations and saturation.

\subsection{JPEG images}
\begin{figure}[t]
    \centering
    \begin{tabular}{cc}
     \begin{subfigure}[b]{0.49\textwidth}
         \centering
         \includegraphics[width=\textwidth]{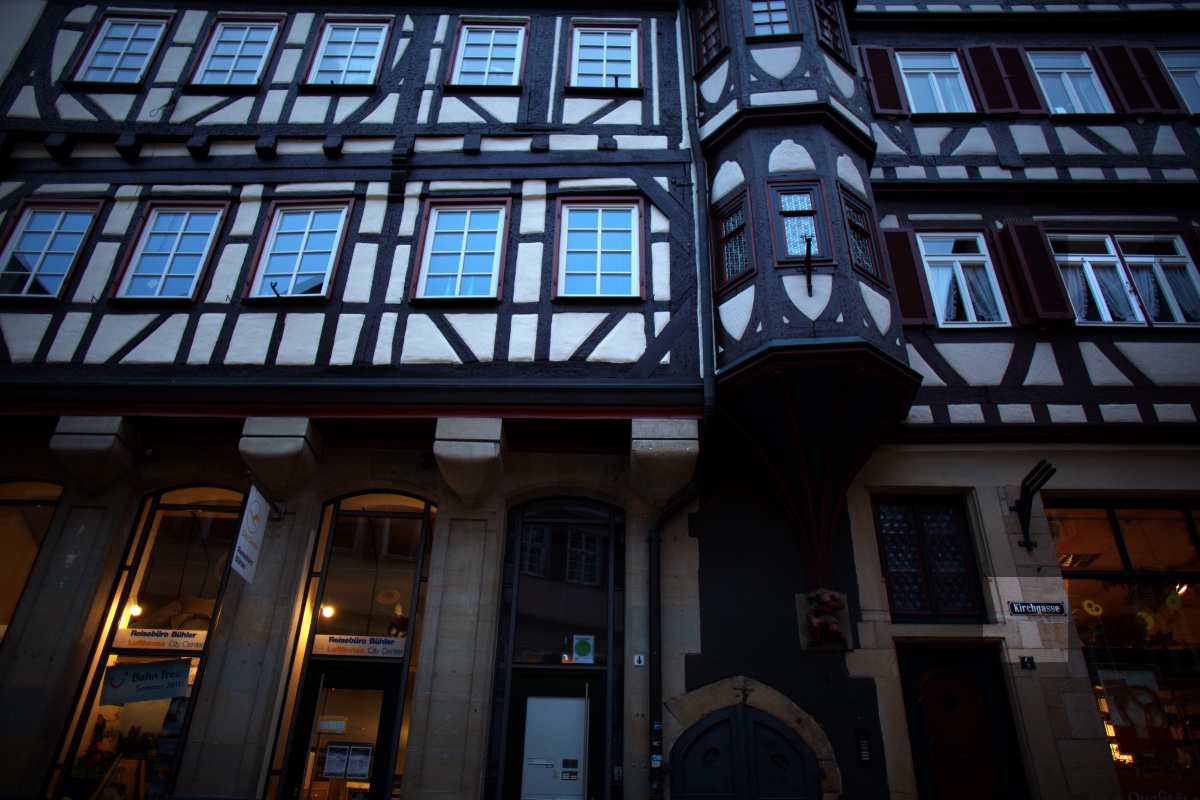}
         \caption{Facade}
     \end{subfigure}  &  
    \begin{subfigure}[b]{0.49\textwidth}
         \centering
         \includegraphics[width=\textwidth]{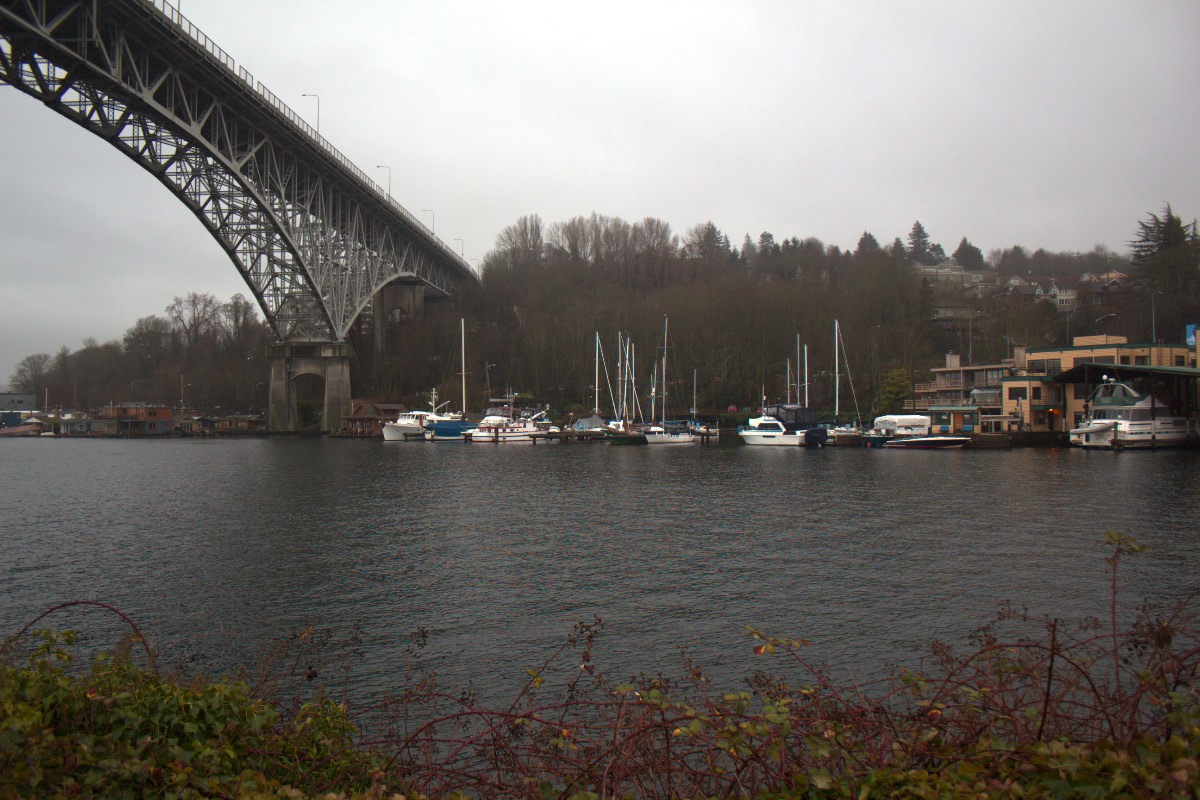}
         \caption{Bridge}
     \end{subfigure} \\
    \end{tabular}
    \caption{The additional JPEG images from \cite{schuler12blind}
    for qualitative evaluation.}
    \label{fig:jpegimages}
\end{figure}

\begin{figure}
    \centering
    \begin{tabular}{ccccc}
     \begin{subfigure}[b]{0.19\textwidth}
         \centering
         \includegraphics[width=\textwidth]{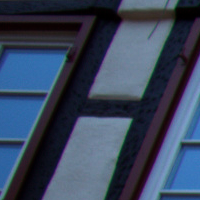}
     \end{subfigure}  &  
     \begin{subfigure}[b]{0.19\textwidth}
         \centering
         \includegraphics[width=\textwidth]{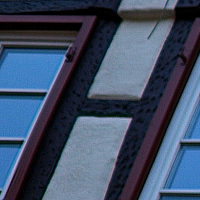}
     \end{subfigure}  & 
     \begin{subfigure}[b]{0.19\textwidth}
         \centering
         \includegraphics[width=\textwidth]{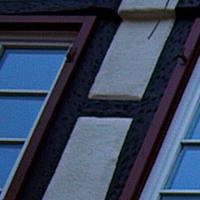}
     \end{subfigure}  & 
     \begin{subfigure}[b]{0.19\textwidth}
         \centering
         \includegraphics[width=\textwidth]{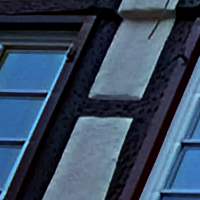}
     \end{subfigure}  & 
    \begin{subfigure}[b]{0.19\textwidth}
         \centering
         \includegraphics[width=\textwidth]{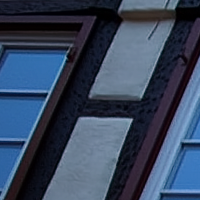}
     \end{subfigure} \\
     \begin{subfigure}[b]{0.19\textwidth}
         \centering
         \includegraphics[width=\textwidth]{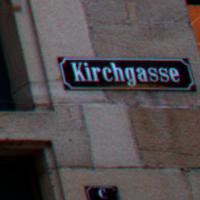}
     \end{subfigure}  &  
     \begin{subfigure}[b]{0.19\textwidth}
         \centering
         \includegraphics[width=\textwidth]{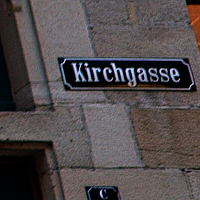}
     \end{subfigure}  & 
     \begin{subfigure}[b]{0.19\textwidth}
         \centering
         \includegraphics[width=\textwidth]{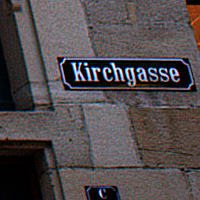}
     \end{subfigure}  & 
     \begin{subfigure}[b]{0.19\textwidth}
         \centering
         \includegraphics[width=\textwidth]{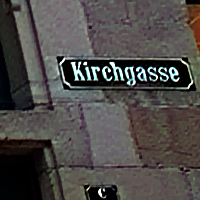}
     \end{subfigure}  & 
    \begin{subfigure}[b]{0.19\textwidth}
         \centering
         \includegraphics[width=\textwidth]{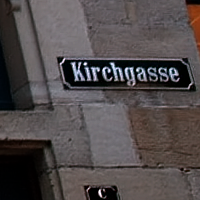}
     \end{subfigure} \\
     \begin{subfigure}[b]{0.19\textwidth}
         \centering
         \includegraphics[width=\textwidth]{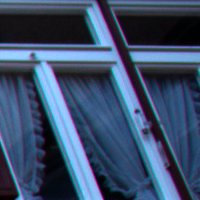}
         \caption{Blurry.}
     \end{subfigure}  &  
     \begin{subfigure}[b]{0.19\textwidth}
         \centering
         \includegraphics[width=\textwidth]{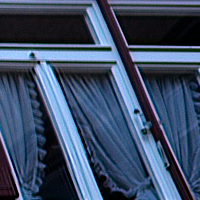}
         \caption{\cite{schuler11nonstationnary}.}
     \end{subfigure}  & 
     \begin{subfigure}[b]{0.19\textwidth}
         \centering
         \includegraphics[width=\textwidth]{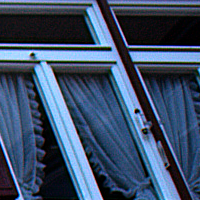}
         \caption{\cite{schuler12blind}.}
     \end{subfigure}  & 
     \begin{subfigure}[b]{0.19\textwidth}
         \centering
         \includegraphics[width=\textwidth]{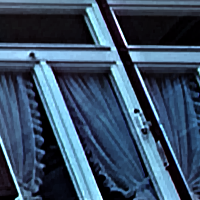}
         \caption{\cite{sun17chrosschannel}.}
     \end{subfigure}  & 
    \begin{subfigure}[b]{0.19\textwidth}
         \centering
         \includegraphics[width=\textwidth]{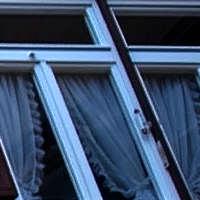}
         \caption{Ours.}
     \end{subfigure} \\
     \end{tabular}
        \caption{Crops for the ``Facade'' image from \cite{schuler12blind}. From left to right: the original blurry image, the non-blind result of Schuler \etal\cite{schuler11nonstationnary}, the blind result of Schuler \etal\cite{schuler12blind}, the blind result of
    Sun \etal\cite{sun17chrosschannel}, and ours.}
    \label{fig:facade}
\end{figure}

\begin{figure}
    \centering
    \begin{tabular}{ccccc}
     \begin{subfigure}[b]{0.19\textwidth}
         \centering
         \includegraphics[width=\textwidth]{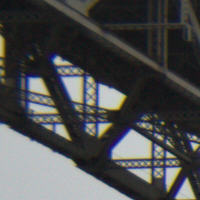}
     \end{subfigure}  &  
     \begin{subfigure}[b]{0.19\textwidth}
         \centering
         \includegraphics[width=\textwidth]{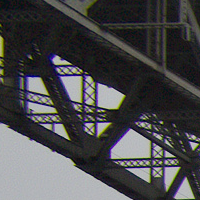}
     \end{subfigure}  & 
     \begin{subfigure}[b]{0.19\textwidth}
         \centering
         \includegraphics[width=\textwidth]{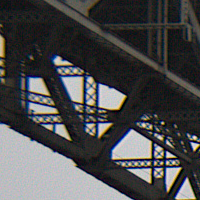}
     \end{subfigure}  & 
     \begin{subfigure}[b]{0.19\textwidth}
         \centering
         \includegraphics[width=\textwidth]{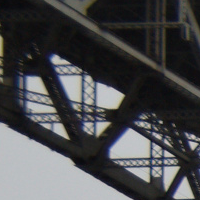}
     \end{subfigure}  & 
    \begin{subfigure}[b]{0.19\textwidth}
         \centering
         \includegraphics[width=\textwidth]{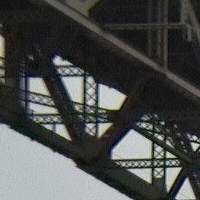}
     \end{subfigure} \\
     \begin{subfigure}[b]{0.19\textwidth}
         \centering
         \includegraphics[width=\textwidth]{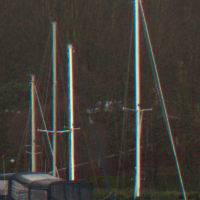}
     \end{subfigure}  &  
     \begin{subfigure}[b]{0.19\textwidth}
         \centering
         \includegraphics[width=\textwidth]{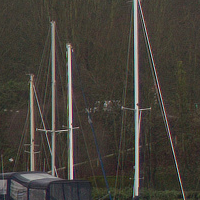}
     \end{subfigure}  & 
     \begin{subfigure}[b]{0.19\textwidth}
         \centering
         \includegraphics[width=\textwidth]{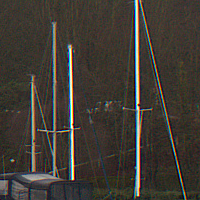}
     \end{subfigure}  & 
     \begin{subfigure}[b]{0.19\textwidth}
         \centering
         \includegraphics[width=\textwidth]{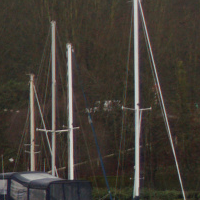}
     \end{subfigure}  & 
    \begin{subfigure}[b]{0.19\textwidth}
         \centering
         \includegraphics[width=\textwidth]{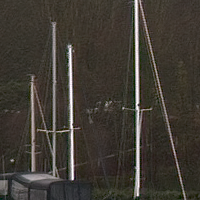}
     \end{subfigure} \\
     \begin{subfigure}[b]{0.19\textwidth}
         \centering
         \includegraphics[width=\textwidth]{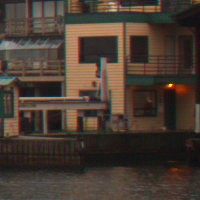}
         \caption{Blurry.}
     \end{subfigure}  &  
     \begin{subfigure}[b]{0.19\textwidth}
         \centering
         \includegraphics[width=\textwidth]{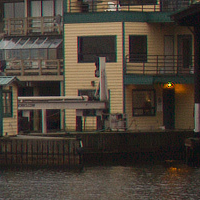}
         \caption{\cite{kee11modeling}.}
     \end{subfigure}  & 
     \begin{subfigure}[b]{0.19\textwidth}
         \centering
         \includegraphics[width=\textwidth]{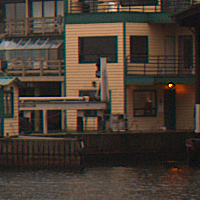}
         \caption{\cite{schuler12blind}.}
     \end{subfigure}  & 
     \begin{subfigure}[b]{0.19\textwidth}
         \centering
         \includegraphics[width=\textwidth]{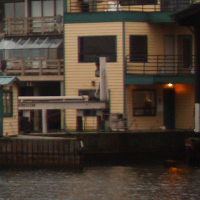}
         \caption{DxO (\cite{schuler12blind}).}
     \end{subfigure}  & 
    \begin{subfigure}[b]{0.19\textwidth}
         \centering
         \includegraphics[width=\textwidth]{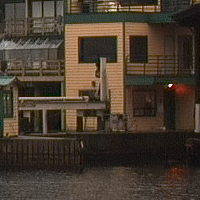}
         \caption{Ours.}
     \end{subfigure} \\
     \end{tabular}
    \caption{Crops for the ``Bridge'' image from \cite{schuler12blind}. From left to right: the original blurry image, the non-blind result of Kee \etal\cite{schuler11nonstationnary}, the blind result of Schuler \etal\cite{schuler12blind}, the blind result of
    DxO (runned by Schuler \etal\cite{schuler12blind}), and ours.}
    \label{fig:bridge}
\end{figure}

We also compare the restoration of JPEG images, with
the methods of \cite{schuler11nonstationnary}, \cite{schuler12blind} and \cite{sun17chrosschannel} when the
images are available. Our method achieves overall the best
results. The images dubbed ``facade'' and
``bridge'' are shown in Figure \ref{fig:jpegimages},
and magnified crops are shown in Figures \ref{fig:facade} and \ref{fig:bridge}.

\end{document}